\documentclass[pra,onecolumn,notitlepage,superscriptaddress,showkeys,showpacs,longbibliography]{revtex4-1}
\usepackage{amsmath,amsfonts,amssymb,amsthm,epsfig,array}
\usepackage{dsfont}
\usepackage{graphics}
\usepackage{float}
\usepackage{verbatim}
\usepackage[dvipsnames]{xcolor}
\usepackage[hidelinks,colorlinks,linkcolor=blue,
citecolor=blue,urlcolor=blue]{hyperref}
\usepackage[titletoc,title]{appendix}
\usepackage{multirow}
\usepackage[format=plain,labelfont=bf]{caption}

\usepackage[normalem]{ulem}

\setcitestyle{number,sort,open={(},close={)}}

\newcommand{\bsl}[1]{\textbf{\text{#1}}}
\newcommand{\shpa}{\shortparallel}

\newcommand{\ii}{\mathrm{i}}
\newcommand{\ee}{\mathrm{e}}

\newcommand{\bra}[1]{\langle #1|}
\newcommand{\ket}[1]{|#1 \rangle}

\newcommand{\Tr}{\mathop{\mathrm{Tr}}}

\newcommand{\SO}{\mathrm{SO}}
\newcommand{\SU}{\mathrm{SU}}

\newcommand{\eqnref}[1]{Eq.\,\eqref{#1}}
\newcommand{\figref}[1]{Fig.\,\ref{#1}}
\newcommand{\tabref}[1]{Tab.\,\ref{#1}}
\newcommand{\secref}[1]{Sec.\,\ref{#1}}
\newcommand{\appref}[1]{Appendix.\,\ref{#1}}
\newcommand{\refcite}[1]{Ref.\,(\onlinecite{#1})}

\newcommand{\mat}[1]{\left(\begin{matrix}#1\end{matrix}\right)}

\newcommand{\eq}[1]{\begin{equation} #1 \end{equation}}

\newcommand{\eqa}[1]{\begin{align}\begin{split} #1 \end{split}\end{align}}
\let\oldAA\AA
\renewcommand{\AA}{\text{\normalfont\oldAA}}
\newcommand{\sgn}[1]{\text{sgn}(#1)}

\newcommand{\ie}{{\emph{i.e.} }}
\newcommand{\eg}{{\emph{e.g.} }}
\newcommand{\TR}{\mathcal{T}}
\newcommand{\cc}{\mathcal{K}}

\newcommand{\eqgammabc}{Eq.\,({\color{blue}3}) }
\newcommand{\eqPETPro}{Eq.\,({\color{blue}2}) }
\newcommand{\eqhpsim}{Eq.\,({\color{blue}5}) }

\newcommand{\tabmainresults}{Tab.\,{\color{blue}1} }
\newcommand{\figGC}{Fig.\,{\color{blue}1} }
\newcommand{\figHgTe}{Fig.\,{\color{blue}2} }

\begin{document}
\title{Piezoelectricity and Topological Quantum Phase Transitions in Two-Dimensional Spin-Orbit Coupled Crystals with Time-Reversal Symmetry}
\author{Jiabin Yu}
\affiliation{Department of Physics, the Pennsylvania State University, University Park, PA, 16802}
\author{Chao-Xing Liu}
\email{cxl56@psu.edu}
\affiliation{Department of Physics, the Pennsylvania State University, University Park, PA, 16802}
\begin{abstract}
Finding new physical responses that signal topological quantum phase transitions is of both theoretical and experimental importance. Here, we demonstrate that the piezoelectric response can change discontinuously across a topological quantum phase transition in two-dimensional time-reversal invariant systems with spin-orbit coupling, thus serving as a direct probe of the transition. We study all gap closing cases for all 7 plane groups that allow non-vanishing piezoelectricity and find that any gap closing with 1 fine-tuning parameter between two gapped states changes either the $Z_2$ invariant or the locally stable valley Chern number. The jump of the piezoelectric response is found to exist for all these transitions, and we propose the HgTe/CdTe quantum well and BaMnSb$_2$ as two potential experimental platforms. Our work provides a general theoretical framework to classify topological quantum phase transitions and reveals their ubiquitous relation to the piezoelectric response.
\end{abstract}
\maketitle

\tableofcontents

\section{Introduction}

The discovery of topological phases and topological phase transitions has revolutionized our understanding of quantum states of matter and quantum phase transitions~\cite{Qi2010TITSC,Hasan2010TI,Chiu2016RMPTopoClas}.
Two topologically distinct gapped phases cannot be adiabatically connected; if the system continuously evolves from one phase to the other, a topological quantum phase transition (TQPT) with the energy gap closing (GC) must occur.
%
A direct way to probe such TQPTs is to detect the discontinuous change of certain physical response functions.
Celebrated examples include the jump of the Hall conductance across the plateau transition in the integer quantum Hall system~\cite{Huckestein1995IQHE,TKNN},
the jump of the two-terminal conductance across the TQPT between the quantum spin Hall (QSH) state and normal insulator (NI) state in a two-dimensional (2D) time-reversal (TR) invariant system~\cite{Bernevig2006BHZ}, and the jump of the magnetoelectric coefficient across the TQPT between the strong topological insulator phase and NI phase in a 3D TR invariant system~\cite{Qi2008TFT,Mogi2017AI,Xiao2018AI,Yu2019MRAI}.
The physical responses in all these examples are induced by the electromagnetic field.
A natural question then arises: can we detect TQPTs with  other types of perturbation?

Here we theoretically answer this question in the affirmative: the discontinuous change of the piezoelectric response is a ubiquitous and direct signature of 2D TQPTs.
The piezoelectric effect, the electric charge response induced by the applied strain, is characterized by the piezoelectric tensor (PET) to the leading order.
PET was originally defined to relate the change of the the charge polarization $\bsl{P}$ with the infinitesimal homogeneous strain, which reads~\cite{Martin1972Piezo}
\eq{
\label{eq:PET_Impro}
\gamma_{ijk}=\left.\frac{\partial P_i}{\partial u_{jk}}\right|_{u_{jk}\rightarrow 0}\ ,
}
where $u_{ij}=(\partial_{x_i} u_j+\partial_{x_j} u_i)/2$ is the strain tensor and $\bsl{u}$ is the displacement at $\bsl{x}$.
The modern theory of polarization~\cite{VKS1993Polarization,KingSmith1993Polarization,
Resta2007ModernPolarization} later identified the above definition as improper~\cite{Vanderbilt2000BPPZ} due to the ambiguity of $\bsl{P}$ in crystals, while the proper definition adds the adiabatic time dependence to $u_{jk}$ and relates it to the bulk current density $J_i$ that can change the surface charge:
\eq{
\label{eq:PET_Pro}
\gamma_{ijk}=
\left.\frac{\partial J_i}{\partial \dot{u}_{jk}}\right|_{u_{jk},\dot{u}_{jk}\rightarrow 0}\ .
}
With \eqnref{eq:PET_Pro}, the PET of an insulating crystal has been derived as~\cite{Vanderbilt2000BPPZ,Wang2018TBPiezo}
\eqa{
\label{eq:gamma_bc}
\gamma_{ijk}=&-e\int \frac{d^2k}{(2\pi)^2} \sum_{n}\left. F_{k_i, u_{jk}}^n\right|_{u_{jk}\rightarrow 0}\ ,
}
where the integral is over the entire first Brillouin zone (1BZ), and $n$ ranges over all occupied bands.
The $F_{k_i, u_{jk}}^n$ term has a Berry-curvature-like expression
\eq{
\label{eq:F}
F_{k_i, u_{jk}}^n=(-\ii)\left[\bra{ \partial_{k_i} \varphi_{n,\bsl{k}}} \partial_{u_{jk}} \varphi_{n,\bsl{k}}\rangle-(k_i\leftrightarrow u_{jk})\right]
}
with $\ket{\varphi_{n,\bsl{k}}}$ the periodic part of the Bloch state in the presence of the strain. (See the Methods for more details.)
The expression indicates an extreme similarity between \eqnref{eq:gamma_bc} and the expression for the Chern number (CN)~\cite{TKNN}.
It is this similarity that motivates us to study the relation between the PET and the TQPT.

Despite the similarity, the topology connected to the PET is essentially different from the CN, since the PET can exist in TR invariant systems whose CNs always vanish.
We, in this work, study the piezoelectric response of 2D TR invariant systems in the presence of the significant spin-orbit coupling (SOC) and demonstrate the jump of all symmetry-allowed PET components across the TQPT.
In particular, we focus on the 7 out of the 17 plane groups (PGs) that allow non-vanishing PET components~\cite{Schwarzenberger1974PG,ITC}, including $p1$, $p1m1$, $c1m1$, $p1g1$, $p3$, $p3m1$, and $p31m$.
The two-fold rotation $C_2$ (with the axis perpendicular to the 2D plane) or the 2D inversion restricts the PET to zero in the other 10 PGs~\cite{Kholkin2008PET}, according to $\gamma_{ijk}= \sum_{i'j'k'}R_{ii'}R_{jj'}R_{kk'}\gamma_{i'j'k'}$ for any $O(2)$ symmetry $R$ of the 2D material.
Through a systematic study, we find that any GC between two gapped states that only requires 1 fine-tuning parameter is a TQPT in the sense that it changes either the $Z_2$ index~\cite{Hasan2010TI,Qi2010TITSC} or the valley CN~\cite{Zhang2013VCN}.
Although the change of the valley CN is locally stable~\cite{Fang2015TCI}, we still treat the corresponding GC as a TQPT, since the two states cannot be adiabatically connected when the valley is well defined.
All the TQPTs contain no stable gapless phase in between two gapped phases, and thereby we refer to them as the {\it direct} TQPTs.
All PET components that are allowed by the crystalline symmetry exhibit discontinuous changes across any of the direct TQPTs, showing the ubiquitous connection.
Interestingly, when the gap closes at momenta that are not TR invariant, the strain tensor $u_{ij}$ acts as a pseudo-gauge field~\cite{Guinea2010PGFGraphene} at the TQPT, making the PET jump directly proportional to the change of the $Z_2$ index or the valley CN.

Our work presents a general framework for the PET jump across the TQPT in 2D TR invariant systems with SOC.
The relation between the PET and the valley CN in the low-energy effective model has been studied in graphene with a staggered potential~\cite{Vaezi2013StrainGraphene}, h-BN~\cite{Droth2016PETBN,Rostami2018PiE}, and monolayer transition metal dichalcogenides (TMDs) XY$_2$ for X=Mo/W and Y=S/Se~\cite{Rostami2018PiE}.
However, these early works have {\it not} pointed out that it is the PET jump (well described within the low-energy effective model) that is the experimental signature directly related to the TQPT, while the PET itself at fixed parameters might contain the non-topological background given by high-energy bands.
Moreover, these works, unlike our systematic study, only considered one specific plane group ($p3m1$) around one specific type of momenta ($K,K'$).
The relation between the PET and the $Z_2$ index were not explored either.
Besides, graphene and h-BN have neglectable SOC, and the TMDs have a large gap, making them not suitable for realizing TQPT.
We thereby propose two realistic material systems, the HgTe/CdTe quantum well (QW) and the layered material BaMnSb$_2$, as potential experimental platforms.
The $Z_2$ TQPT and PET jump can be achieved by varying the thickness or the gate voltages in the HgTe/CdTe QW or by tuning lattice distortion in BaMnSb$_2$.

%
%

\section{Results}

\subsection{PET jump across a Direct QSH-NI TQPT}
\label{sec:p1}

\begin{figure}[h]
    \centering
    \includegraphics[width=0.9\columnwidth]{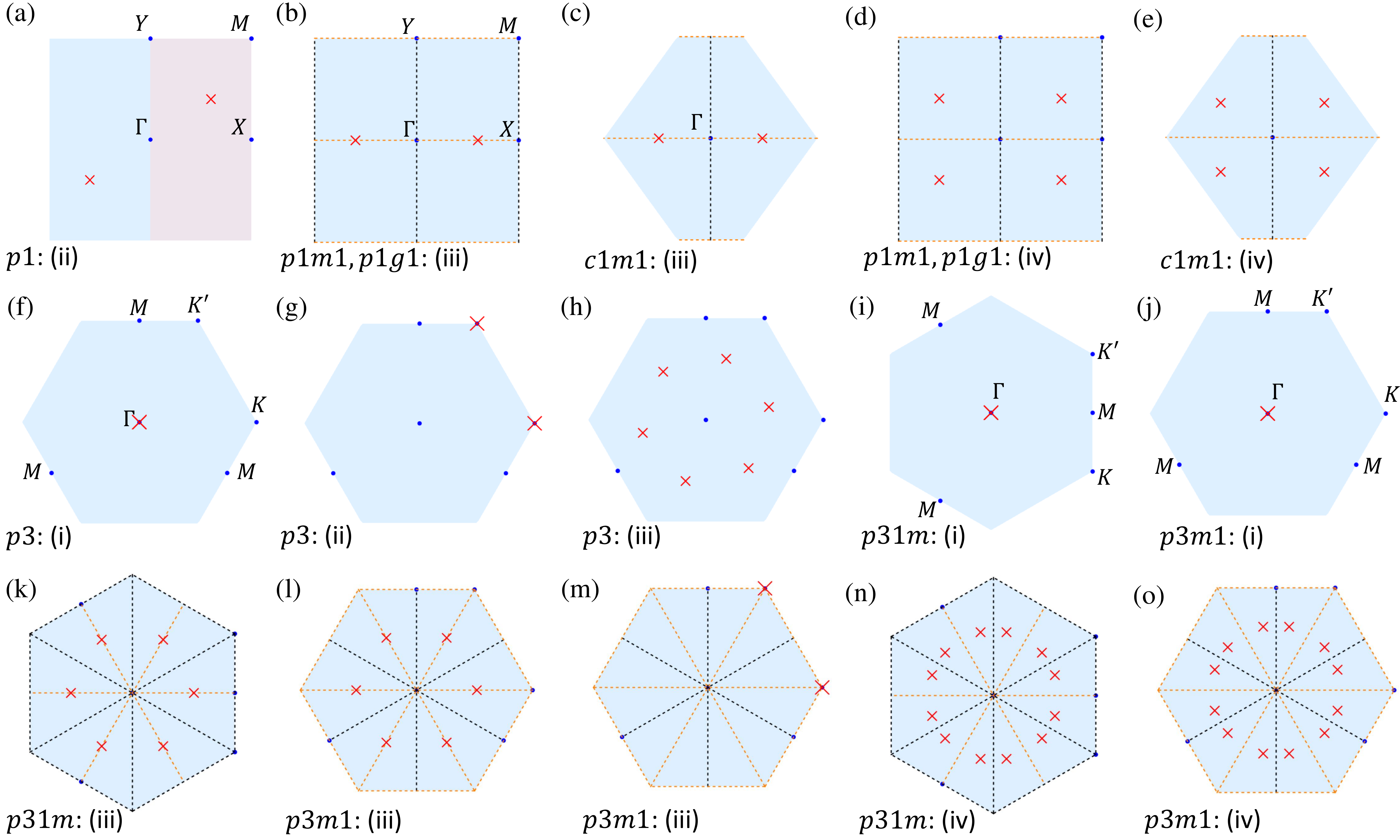}
    \caption{\textbf{GC cases with 1 fine-tuning parameter.} The figure shows the GC cases between insulating states with 1 fine-tuning parameter for all 7 PGs with non-vanishing PET.
    The red cross labels the GC momenta, the light blue background indicates the 1BZ, and the light red part in (a) indicates the half 1BZ.
    The black and orange dashed lines label the momenta invariant under the mirror/glide symmetry and the combination of mirror/glide and TR symmetries, respectively.
    The figures are first grouped according to the PGs and then ordered based on the GC scenarios listed in \tabref{tab:main_results}, whose labels are next to the names of the PGs.
    }
    \label{fig:GC}
\end{figure}

We start from a simple example of the TQPT discussed in \refcite{Murakami2007QSH}.
They (in the example of our interest) considered the case with no crystalline symmetries other than the lattice translation (PG $p1$) and focused on the GC at two momenta $\pm \bsl{k}_0$ that are not TR invariant momenta (TRIM), as labeled by red crosses in \figref{fig:GC}(a).
The low-energy effective theory for the electron around $\bsl{k}_0$ can be described by the Hamiltonian of a 2D massive Dirac fermion~\cite{Murakami2007QSH}
\eqa{
\label{eq:h_+_0_p1_sim}
 h_{+,0}(\bsl{q})= E_0(\bsl{q})\sigma_0+v_x q_1\sigma_x+v_y q_2\sigma_y + m \sigma_z\ ,
}
where $\bsl{q}=\bsl{k}-\bsl{k}_0$, $m$ is the tuning parameter for the TQPT, and $\sigma$'s are Pauli matrices.
In the above Hamiltonian, the unitary transformation on the bases and the scaling/rotation of $\bsl{q}$ are performed for the simplicity of the Hamiltonian; the latter generally makes $q_1,q_2$ along two non-orthogonal directions. (See {Appendix C} for details.)
The effective Hamiltonian at $-\bsl{k}_0$ is related to $h_{+,0}$ by the TR symmetry.
After choosing appropriate bases at $-\bsl{k}_0$, the TR symmetry can be represented as $\TR\dot{=} \ii\sigma_y \cc$ with $\cc$ the complex conjugate, leading to
\eqa{
\label{eq:h_-_0_p1_sim}
 h_{-,0}(\bsl{q})=  E_0(-\bsl{q})\sigma_0+v_x q_1\sigma_x+v_y q_2\sigma_y - m \sigma_z \ .
}

According to \refcite{Murakami2007QSH}, the TQPT between the QSH insulator and the NI (distinguished by the $Z_2$ index) occurs when the mass $m$ in $h_{\pm,0}(\bsl{q})$ changes its sign.
The argument used to determine change of the $Z_2$ index was presented in \refcite{Moore20072DTRI} and is discussed below for integrity.
Since there is no inversion symmetry in PG $p1$, the $Z_2$ index can be determined from the CN of the contracted half first Brillouin zone (1BZ), where the half 1BZ is chosen such that its Kramers' partner covers the other half.
Specifically, the $Z_2$ index is changed (unchanged) by the GC if the CN of the contracted half 1BZ changes by an odd (even) integer.
Without loss of generality, let us choose the half 1BZ to contain $\bsl{k}_0$,  as shown in \figref{fig:GC}(a).
Since $h_{+,0}$ is a 2D gapped Dirac Hamiltonian, the CN of the contracted half 1BZ changes by $\Delta N_+=-\sgn{v_x v_y}$ as $m$ increases from $0^-$ to $0^+$, featuring a direct QSH-NI TQPT as $v_x v_y$ is typically nonzero.

We next discuss the piezoelectric effect in this simple effective model.
To do so, we need to introduce the electron-strain coupling around $\pm\bsl{k}_0$ based on the TR symmetry:
\eq{
\label{eq:h_+-_1_p1}
h_{\pm,1}(u)=\xi_{0,ij}\sigma_0 u_{ij}\pm \xi_{a',ij}\sigma_{a'} u_{ij}\ ,
}
where the duplicated indexes, including $a'=x,y,z$ and $i,j=1,2$, are summed over henceforth unless specified otherwise.
$\xi$'s are the material-dependent coupling constants between the low-energy electrons and the strain tensor, which obey $\xi_{a,ij}=\xi_{a,ji}$ with $a=0,x,y,z$ owing to $u_{ij}=u_{ji}$ and are related to the electron-phonon coupling~\cite{Suzuura2002Phonon-Electron}.
The full form of the effective Hamiltonian is then given by
\eq{
\label{eq:h_+-_p1_sim}
h_\pm(\bsl{q},u)=h_{\pm,0}(\bsl{q})+h_{\pm,1}(u)\ .
}

To use \eqnref{eq:gamma_bc}, we simplify \eqnref{eq:h_+-_p1_sim} by neglecting the $E_0$ term, which has no influence on the piezoelectric response of insulators (see {Appendix A}).
When $\xi_{x,ij}=\xi_{y,ij}=0$, the Hamiltonian $h_\pm$ has effective inversion symmetry within each valley, $\sigma_z h_{\pm}(-\bsl{q},u) \sigma_z=h_{\pm}(\bsl{q},u)$, which forbids the piezoelectric effect.
Thus, $\xi_{0,ij}$ and $\xi_{z,ij}$ terms cannot contribute to the PET, and neglecting them leads to a further simplified version of \eqnref{eq:h_+-_p1_sim}:
\eqa{
\label{eq:A_pse_p1}
h_{\pm}(\bsl{q},u)=&\left[v_x (q_1\pm A_1^{pse})\right]\sigma_x+\left[v_y (q_2\pm A_2^{pse})\right]\sigma_y\\
& \pm m \sigma_z\ ,
}
where $A_1^{pse}=\xi_{x,ij}u_{ij}/v_x$ and $A_2^{pse}= \xi_{y,ij}u_{ij}/v_y$.
The above form suggests that the remaining strain terms, $\xi_{x,ij}$ and $\xi_{y,ij}$, serve as the pseudo-gauge field $A_i^{pse}$ that has opposite signs for two valleys $\pm \bsl{k}_0$~\cite{Guinea2010PGFGraphene,Guinea2010StrainQHEGraphene,
Rostami2018PiE,Yu2019MRAI}.
As the strain tensor only exists in the form of $q_i\pm A_i^{pse}$, the derivative with respect to $u_{ij}$ in \eqnref{eq:gamma_bc} can be replaced by the derivative with respect to the momentum as
\eq{
\partial_{u_{ij}} \ket{\varphi_{\pm,\bsl{q}}}=\frac{\partial A_{i'}^{pse}}{\partial u_{ij}}\partial_{A_{i'}^{pse}} \ket{\varphi_{\pm,\bsl{q}}}=\pm \frac{\partial A_{i'}^{pse}}{\partial u_{ij}}\partial_{q_{i'}} \ket{\varphi_{\pm,\bsl{q}}}\ ,
}
where $\varphi_\pm$ are the occupied bands of $h_{\pm}$.
Substituting the above equation into \eqnref{eq:gamma_bc} leads to
\eqa{
\label{eq:PET_F}
&\gamma_{1ij}^{eff}=-e\int \frac{d^2q}{(2\pi)^2} \sum_{\alpha=\pm} \alpha F^\alpha_{12}(\bsl{q})  \frac{\partial A_{2}^{pse}}{\partial u_{ij}} \\
&\gamma_{2ij}^{eff}=e\int \frac{d^2q}{(2\pi)^2} \sum_{\alpha=\pm} \alpha F^\alpha_{12}(\bsl{q})  \frac{\partial A_{1}^{pse}}{\partial u_{ij}}\ ,
}
where $F_{12}^\pm(\bsl{q})$ is the conventional Berry curvature of the occupied band of $h_\pm(\bsl{q},0)$.
The superscript \emph{eff} means that we neglect the contribution from bands beyond the effective model \eqnref{eq:h_+-_p1_sim}, indicating that the above equation is not the complete PET.
Nevertheless, it can accurately give the PET change across the TQPT since high-energy bands experience an adiabatic deformation and the corresponding background PET contribution should remain unchanged at the transition $(m=0)$.
As $m$ varies from $0^-$ to $0^+$, \eqnref{eq:PET_F} gives the change of PET $\Delta \gamma_{ijk}$ as
\eqa{
\label{eq:PET_jump_p1}
&\Delta\gamma_{1ij}=-e\frac{\Delta N_+}{\pi} \frac{\xi_{y,ij}}{v_y} \\
&\Delta\gamma_{2ij}=e\frac{\Delta N_+}{\pi} \frac{\xi_{x,ij}}{v_x} \ .
}
The PET jump shown in the above equation is nonzero since $v_x v_y$ and the electron-strain coupling $\xi$'s are typically non-zero.
We thus conclude that for $p1$ group, a jump of PET that is directly proportional to the change of the $Z_2$ index occurs across the TQPT, when the gap closes not at TRIM.

The PET jump can be physically understood based on \eqnref{eq:PET_Pro}.
Let first focus on one GC momentum, say $\bsl{k}_0$.
Since the strain tensor couples to the electron in the way similar to the $U(1)$ gauge field as shown in \eqnref{eq:h_+-_p1_sim}, $\dot{u}_{jk}$ should act like a electric field on the electron.
According to \eqnref{eq:PET_Pro}, $\gamma_{ijk}$ should then behave like the Hall conductance, whose jump is proportional to the change of CN $\Delta N_+$.
Now we include the other GC momentum $-\bsl{k}_0$.
Unlike the actual $U(1)$ gauge field, the pseudo-gauge field given by the strain couples oppositely to the electron at the two GC momenta (\eqnref{eq:h_+-_p1_sim}).
The opposite signs of the coupling can cancel the opposite signs of the Berry curvature, and thus, in contrast to the actual Hall conductance, the contributions to $\gamma_{ijk}$ from $\pm\bsl{k}_0$ add up to a nonzero value instead of canceling each other, leading to the non-zero topological jump in \eqnref{eq:PET_jump_p1}.

\subsection{Classification of Direct 2D TQPTs and PET jumps for 7 PGs}
\label{sec:PGs}
The above section discusses an example of 2D QSH-NI TQPT for the $p1$ PG and illustrates the main picture of the relation between the 2D TQPT and the PET jump.
It is well-known that the crystalline symmetry imposes strong constraints on the PET~\cite{Kholkin2008PET} (see the Methods).
%
Topological states in different space/plane groups have been classified based on the topological quantum chemistry~\cite{Bradlyn2017TQC,Bradlyn2018TQCGraph,
Cano2018TQCEBR,Cano2018DisEBR,Bradlyn2019Fragile,Wieder2018AXIFragile,
Wieder2019DSMFrigle}, the symmetry indicator~\cite{Po2017SymIndi,Kruthoff2017TCI,
Watanabe2018SIMSG,Song2018SI}, and other early methods~\cite{Dong2016TCIRep,Chiu2013ReflectionTITSC,Shiozeki2014TCITCSC}.
On the contrary, only a small number of works~\cite{Murakami2007QSH,Ahn2017C2T,Kruthoff2017TCI,Yang20172DGC} have studied the crystal symmetry constraint on the GC forms of the TQPTs.
While the GC between non-degenerate states was studied in \refcite{Yang20172DGC} for various layer groups in the presence of TR symmetry and SOC, the GC that involves degenerate states, like between two Kramers' pairs, has not been explored.
In particular, the topology change and the PET jump across any GC case with codimension 1 have not been discussed.
As the substrate, on which the 2D materials are grown, typically reduces layer groups to PGs by breaking the extra symmetries, a study based on PG is typically enough for experimental predictions.
Therefore, we next present a comprehensive study on the GC forms of TQPTs in all 7 PGs that allow nonvanishing PET, namely $p1$, $p1m1$, $c1m1$, $p1g1$, $p3$, $p31m$, and $p3m1$. The main results are summarized in \figref{fig:GC} and \tabref{tab:main_results}, as discussed below.
The other 10 PGs ($p2$, $p2mm$, $p2mg$, $p2gg$, $c2mm$, $p4$, $p4mm$, $p4gm$, $p6$, and $p6mm$) have vanishing PET due to the existence of inversion symmetry or $C_2$ rotation symmetry, and are briefly discussed in {Appendix B}.

TQPTs in different PGs can be analyzed in the following three steps.
%
In the first step, we classify the GC based on the GC momenta and the symmetry property of the bands involved in the GC.
To do so, we define the group $\mathcal{G}_0$ for a GC momentum $\bsl{k}_0$ such that $\mathcal{G}_0$ contains all symmetry operations that leave $\bsl{k}_0$ invariant (including the little group of $\bsl{k}_0$ and TR-related operations).
We start with a coarse classification based on $\mathcal{G}_0$, which leads to 2 scenarios for $p1$, 3 scenarios for $p3$, and 4 scenarios for $p1m1$, $c1m1$, $p1g1$, $p31m$, and $p3m1$, as listed in \tabref{tab:main_results} and the Methods.
%
%
To illustrate this classification, we consider the $p3$ group as an example, which contains 3 different scenarios.
In scenario (i), the GC is located at TRIM ($\TR\in \mathcal{G}_0$), \ie the $\Gamma$ point or three $M$ points in \figref{fig:GC}(f).
In scenario (ii), the GC occurs simultaneously at $K$ and $K'$ where $\mathcal{G}_0$ contains $C_3$ but no $\TR$ (\figref{fig:GC}(g)).
%
In scenario (iii), the GC occurs at six generic momenta ($\mathcal{G}_0$ only contains lattice translations) that are related by $C_3$ rotation and TR (\figref{fig:GC}(h)).
The classification of GC momenta is coarse here since $\mathcal{G}_0$ can still vary within one scenario.
For example, in scenario (i) of $p3$, $\mathcal{G}_0$ at $\Gamma$ contains $C_3$ while $\mathcal{G}_0$ at $M$ does not.
Moreover, even at a certain GC momentum with a certain $\mathcal{G}_0$, the symmetry properties of bands involved in the GC may vary.
For example, at $K$ in scenario (ii) of $p3$, the gap may close between two states with the same or different $C_3$ eigenvalues.
Therefore, we further refine our classification by taking these subtleties into consideration and classify each GC scenario into finer GC cases.

In the second step, for each GC case, we construct a symmetry-allowed low-energy effective Hamiltonian that well captures the GC and count the number of fine-tuning parameters.
Since $\mathcal{G}_0$ and the symmetry properties of the bands involved in the GC are fixed in one GC case, the form of the effective Hamiltonian can be unambiguously determined. (See details in {Appendix B and C}.)
After obtaining the effective Hamiltonian, we can count the number of fine-tuning parameters required for each GC and select out all GC cases that require only 1 fine-tuning parameter (or equivalently has codimension 1), as shown in \figref{fig:GC}.
Only these cases can be direct TQPTs between two gapped phases, since any two gapped states in the parameter space are adiabatically connected if 2 or more fine-tuning parameters are required to close the gap, and 0 codimension means there is a stable gapless phase in between two gapped phases.
Our analysis shows that all GC cases in scenarios (i) for $p1$, (i) and (ii) for $p1m1$, $c1m1$, and $p1g1$, and (ii) for $p3m1$ and $p31m$ need $0$ fine-tuning parameter or more than 1 fine-tuning parameters and thus cannot correspond to the direct TQPTs, while codimension-1 GC cases can exist in all other scenarios.

In the third and final step, we demonstrate the topological nature of all the codimension-1 GC cases by evaluating the change of certain topological invariants and derive the corresponding PET jump.
As shown in \tabref{tab:main_results}, the $Z_2$ index is changed in all codimension-1 GC cases of scenarios (ii) for $p1$, (iii) for $p1m1$, $c1m1$, and $p1g1$, (i)-(iii) for $p3$, and (i) and (iii) for $p3m1$ and $p31m$, while the valley CN is changed for all codimension-1 GC cases of the scenarios (iv) for $p1m1$, $c1m1$, $p1g1$, $p3m1$, and $p31m$.
We would like to emphasize that although valley CN itself is in general not quantized in a gapped phase, the change of valley CN across a gap closing is quantized and has physical consequence~\cite{Martin2010VCN}. (See the Methods for more details.)
According to \figref{fig:GC}, the $Z_2$ cases either close the gap at TRIM or have an odd number of Dirac cones in half 1BZ, while all the valley CN cases (\figref{fig:GC}(d-e) and \figref{fig:GC}(n-o)) have an even number of Dirac cones in half 1BZ, forbidding the change of the $Z_2$ index.
Nevertheless, no matter which type, they all lead to discontinuous changes of the symmetry-allowed PET components. (See detailed calculation of PET in {Appendix B}.)

In sum, we conclude that for all 7 PGs with non-vanishing PET, all the GC cases between two gapped phases with 1 fine-tuning parameter are direct TQPTs that change either $Z_2$ index or valley CN, and they all induce the discontinuous change of the symmetry-allowed PET components.
Based on these results, we propose the following criteria to find realistic systems to test our theoretical predictions: (i) whether it breaks the 2D inversion or two-fold rotation with axis perpendicular to the 2D plane, (ii) whether it has significant SOC, and (iii) whether there is a tunable way to realize the GC. 
Applying these conditions to the existing material systems for 2D TQPT, we find two realistic material systems, namely the HgTe/CdTe QW and the layered material BaMnSb$_2$, which are studied in the following.

\begin{table}[h]
    \centering
    \begin{tabular}{|c||c|c||c|c|c|c||c|c|c||c|c|c|c|}
    \hline
        PGs & \multicolumn{2}{c||}{$p1$} & \multicolumn{4}{c||}{$p1m1$, $c1m1$, $p1g1$} & \multicolumn{3}{c||}{$p_3$} & \multicolumn{4}{c|}{$p3m1$,$p31m$}\\
    \hline
        Scenario & (i) & (ii) & (i) & (ii) & (iii) & (iv)& (i) & (ii) & (iii) & (i) & (ii) & (iii) & (iv)\\
    \hline
        Codim-1 GC & $\times$ & (a) & $\times$ & $\times$ & (b-c)& (d-e) & (f) & (g) & (h) &  (i-j) & $\times$ &(k-m) & (n-o)\\
    \hline
        Topo. Inv. & N/A & $Z_2$ & N/A & N/A & $Z_2$ & VCN & $Z_2$ & $Z_2$ & $Z_2$ & $Z_2$ & N/A & $Z_2$ & VCN\\
    \hline
       PET Jump  & N/A & $\checkmark$ & N/A & N/A & $\checkmark$ & $\checkmark$ & $\checkmark$ & $\checkmark$ & $\checkmark$ & $\checkmark$ & N/A & $\checkmark$ & $\checkmark$\\
    \hline
    \end{tabular}
    \caption{\textbf{Summary for all 7 PGs with non-vanishing PET.}
    The scenarios are classified by the symmetries that leave the GC momenta invariant, as shown in the Methods.
    Codim-1 GC means the GC cases with 1 fine-tuning parameter or codimension 1.
    If at least one GC case between gapped states with 1 fine-tuning parameter exists in the corresponding scenario, the subfigures in \figref{fig:GC} that illustrate the GC momenta are referred to; otherwise, we fill in a $\times$.
    Topo. Inv. labels the topological invariant changed by the GC, $Z_2$ means the $Z_2$ index, and VCN means the corresponding case changes the valley CN when the valley is well-defined.}
 \label{tab:main_results}
\end{table}

\subsection{HgTe/CdTe Quantum Well}
\label{sec:HgTe}
It has been demonstrated~\cite{Bernevig2006BHZ,Konig2007QSHHgTe} that the TQPT between the QSH insulator and NI phases in the HgTe/CdTe QW can be achieved by tuning the HgTe thickness $d$.
Tuning applied electric field $\mathcal{E}$ was theoretically predicted as an alternative way to achieve TQPT~\cite{Li2009HgTeEfield,Rothe2010HgTe}, making the system an ideal platform to study the PET jump at TQPTs.
Here, the stacking direction of the QW is chosen to be (111) instead of the well-studied (001) direction~\cite{Novik2005HgMnTe}, since the latter would allow a two-fold rotation that forbids PET.
Without the applied electric field, the (111) QW has the TR symmetry and the $C_{3v}$ symmetries (generated by three-fold rotation along $(111)$ and the mirror perpendicular to $(\bar{1}10)$); adding electric field along $(111)$ does not change the symmetry properties.
We should then expect one independent symmetry-allowed PET component $\gamma_{222}$ similar to \eqnref{eq:PET_sym_p3m} in the Methods, where 2 labels the direction $(11\bar{2})$.

The electronic band structure of the (111) QW can be described by the 6-band Kane model with the bases $(\ket{\Gamma_6,\pm\frac{1}{2}}$, $\ket{\Gamma_8,\pm\frac{3}{2}}$, $\ket{\Gamma_8,\pm\frac{1}{2}})$.
The electric field $\mathcal{E}$ along (111) can be introduced by adding a linear electric potential that is independent of orbitals and spins.
In this electron Hamiltonian, there are two inversion-breaking (IB) effects, the inherent IB effect in the Kane model and the applied electric field, and we neglect the former for simplicity.
Note that such approximation does not lead to vanishing PET even for $\mathcal{E}=0$ because the IB electron-strain coupling will be kept.

\begin{figure}[h]
\includegraphics[width=0.6\columnwidth]{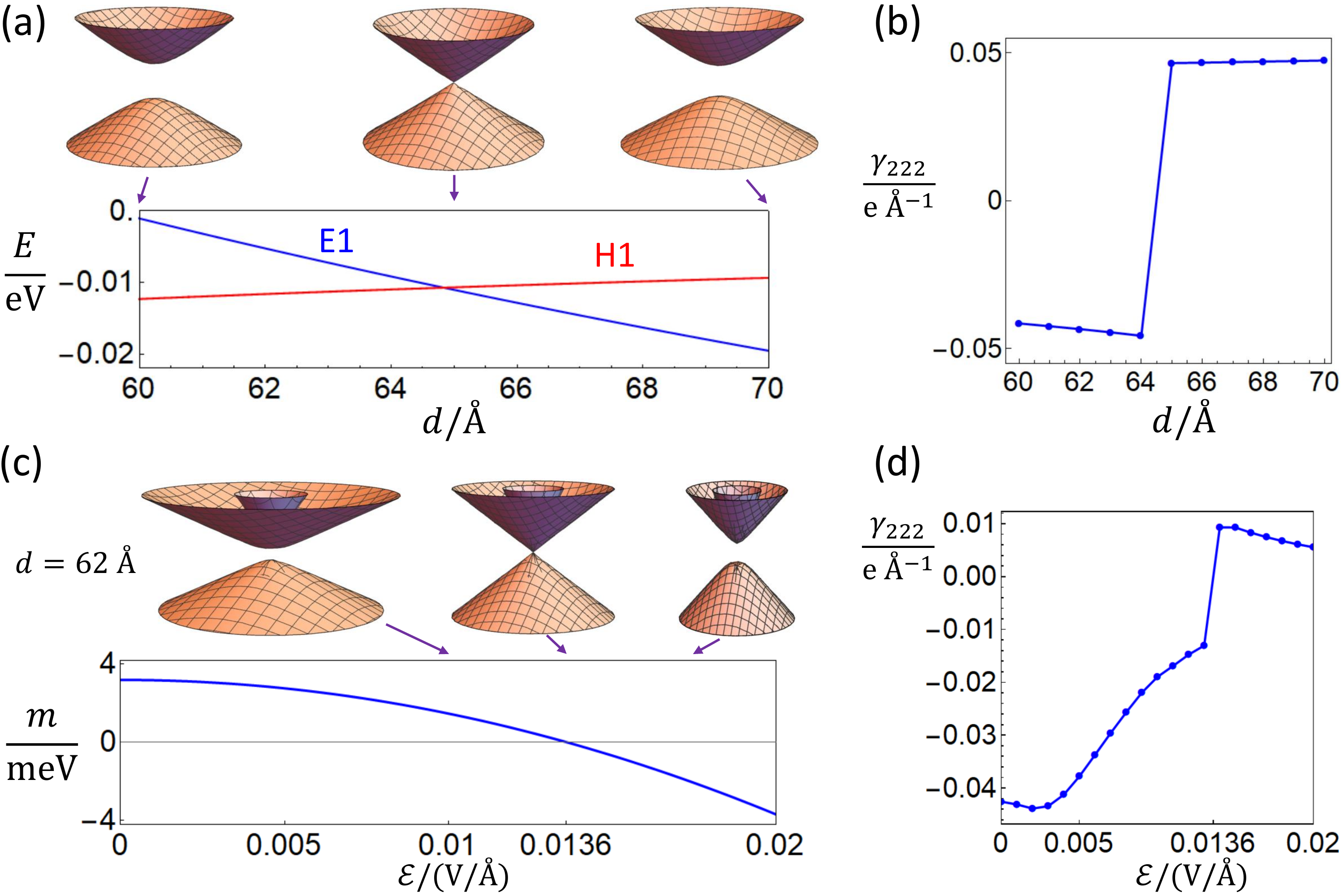}
\caption{\textbf{HgTe/CdTe QW.} This figure shows the energy dispersion and the PET of the HgTe QW with the stacking direction $(111)$.
In (\textbf{a}), the lower panel shows the energy of E1 (blue) and H1 (red) bands at $\Gamma$ point as a function of the width $d$, and the upper panel shows the energy dispersion at $d=60,65,70\AA$ from left to right, respectively.
The GC happens around $d\approx 65\AA$, which is slightly different form the well-known $d=63\AA$ reported in \refcite{Konig2007QSHHgTe} for the $(001)$ stacking direction owing to the anisotropy effect.
(\textbf{b}) shows the PET component $\gamma_{222}$ as a function of $d$.
In (\textbf{c}), the lower panel plots gap $m$ as a function of the electric field $\mathcal{E}$ with $d=62$\AA, showing that the gap closes at $\mathcal{E}\approx 0.0136${V$\AA^{-1}$}.
The upper panel of (c) demonstrates the energy dispersion at $\mathcal{E}=0.01,0.0136,0.017${V$\AA^{-1}$} from left to right, respectively.
(\textbf{d}) shows the PET component $\gamma_{222}$ as a function of $\mathcal{E}$.
}
\label{fig:HgTe}
\end{figure}

We first discuss the inversion-invariant $\mathcal{E}=0$ case and focus on the PET jump induced by varying the width $d$.
In this case, there are two double degenerate bands closest to the Fermi energy, namely $\ket{E_1,\pm}$ and $\ket{H_1,\pm}$ bands with opposite parities.
With the method proposed in \refcite{Bernevig2006BHZ}, we find that the gap between two bands closes at the $\Gamma$ point around $d=65\AA$ as shown in \figref{fig:HgTe}(a).
The GC must be a $Z_2$ TQPT owing to the opposite parities of the two bands, and it belongs to scenario (i) of $p3m1/p31m$ discussed in \tabref{tab:main_results} and the Methods.
We further include the electron-strain coupling, and numerically plot the independent PET component $\gamma_{222}$ as the function of the width in \figref{fig:HgTe}(b), which shows a jump around $d=65\AA$. (See {Appendix E}.)

Next we study the TQPT induced by the applied electric field.
In order to realize the GC at a nonzero value of the electric field, we fix the width of the QW at $d=62\AA$, away from $65\AA$.
After adding the linear electric potential along (111) in the 6-band Kane model, we numerically find that the GC at $\Gamma$ point happens at $\mathcal{E}\approx 0.0136$V {$\AA^{-1}$}, as shown in \figref{fig:HgTe}(c).
Such GC belongs to scenario (i) of $p3m1/p31m$ and is still a $Z_2$ TQPT since the extra IB term cannot influence the $Z_2$ topology change.
The PET component $\gamma_{222}$ is numerically shown in \figref{fig:HgTe}(d), showing the jump across the TQPT.
The PET jump in \figref{fig:HgTe}(b) and (d) has the order $10\sim 100$pC{ m$^{-1}$}, and thus is possible to be probed by the current experimental technique~\cite{Zhu2014PET}.

\begin{figure}[h]
\includegraphics[width=0.6\columnwidth]{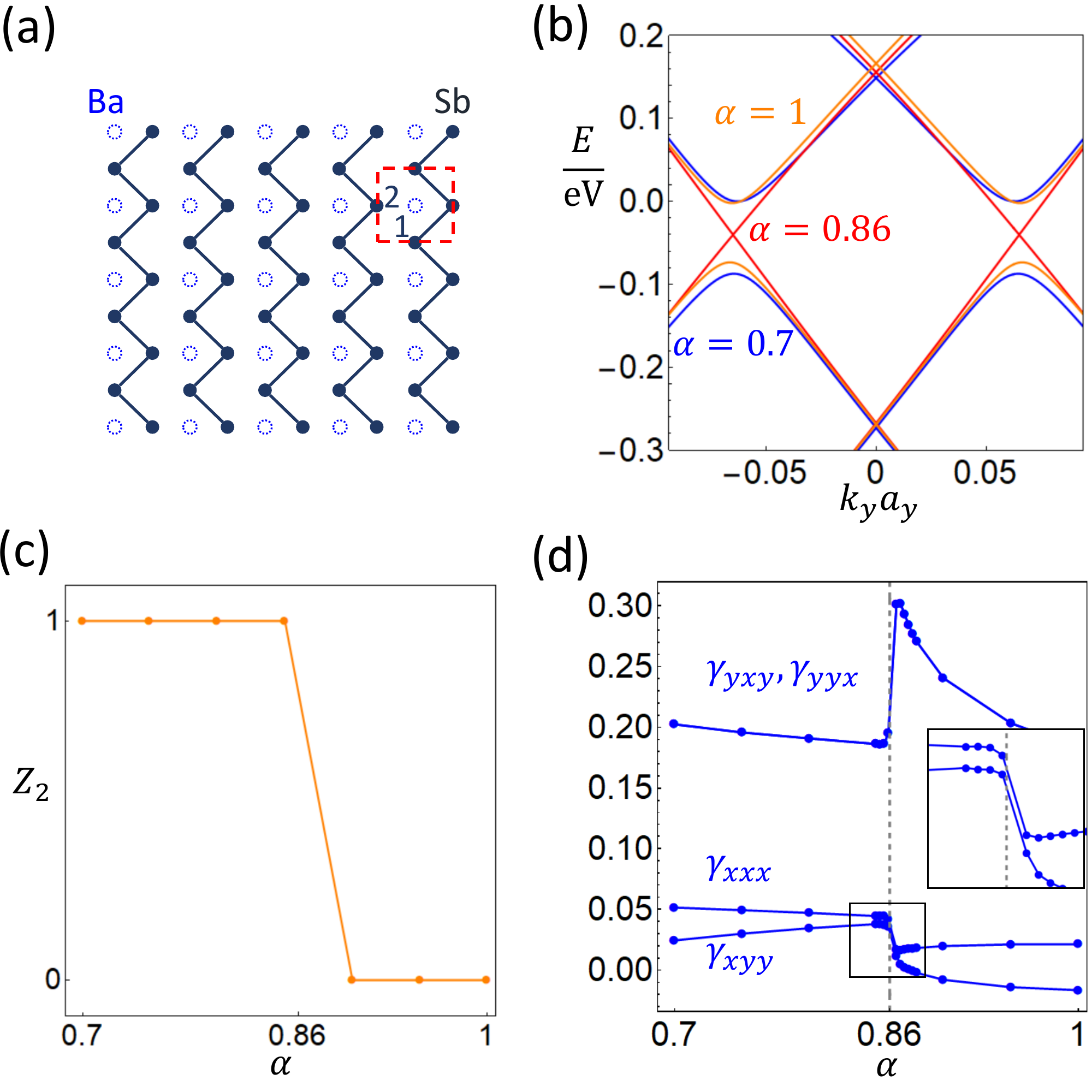}
\caption{\textbf{Layered Material BaMnSb$_2$.}
(\textbf{a}) illustrates the Ba-Sb layer, where each dashed  circle stands for the projection of two Ba atoms onto the Sb layer and the solid dots are Sb atoms.
The solid lines connecting Sb atoms indicate the zig-zag distortion, and the red dashed box marks the unit cell with $1$ and $2$ labeling the two Sb atoms.
(\textbf{b}) The band structure of the TB model for BaMnSb$_2$ along $M-X-M$ for $\alpha=0.86$ (red), $\alpha=1$ (orange), and $\alpha=0.7$ (blue), respectively, where $X$ is at $k_y=0$.
(\textbf{c}) and (\textbf{d}) plot the $Z_2$ index and PET components obtained from the TB model as a function of $\alpha$, respectively.
In (d), the PET components are in the unit {e$\AA^{-1}$}, the gray dashed line is at $\alpha=0.86$, and the inset is the zoom-in version of the boxed region.
}
\label{fig:bms}
\end{figure}

\subsection{Layered Material BaMnSb$_2$}

BaMnSb$_2$ is a 3D layered material that consists of Ba-Sb layers and Mn-Sb layers, which are stacked alternatively along the $(001)$ direction (or equivalently $z$ direction).
The electrons in $p_x$ and $p_y$ orbitals of Sb atoms in the Ba-Sb layers account for the transport of the material.
Owing to the insulating Mn-Sb layers, the tunneling along the $z$ direction among different Ba-Sb layers is much weaker than the in-plane hopping terms, and thus BaMnSb$_2$ can be treated as a quasi-2D material~\cite{Liu2019SCM3DQHE}.
Therefore, we can only consider one Ba-Sb layer, whose structure is shown in \figref{fig:bms}(a).
Owing to the zig-zag distortion of the Sb atoms (solid lines in \figref{fig:bms}(a)), the symmetry group that captures the main physics is spanned by the TR symmetry $\TR$ and two mirror operations $m_y$ and $m_z$ that are perpendicular to $y$ and $z$ axes, respectively.
The mirror symmetry $m_z$ does nothing but guarantee the z-component of the spin to be a good quantum number, allowing us to view the system as a spin-conserved TR-invariant 2D system with PG $p1m1$.
Slightly different from the demonstration in the Methods, the mirror here is perpendicular to $y$ instead of $x$, and thereby PG $p1m1$ now requires $\gamma_{yyy}=\gamma_{yxx}=\gamma_{xyx}=\gamma_{xxy}=0$ and leaves the other four components as symmetry-allowed.

To describe this system, a tight-binding model with $p_x$ and $p_y$ orbitals of Sb atoms was constructed in \refcite{Liu2019SCM3DQHE} based on the first-principle calculation, and the form of the model {is} reviewed in {Appendix F} for integrity.
This model qualitatively captures all the main features of the electronic band structure of BaMnSb$_2$.
The key parameter of the model is the distortion parameter $\alpha$ that describes the zig-zag distortion of the Sb atoms.
When $\alpha$ is tuned to a critical value {$\alpha_{\text{c}}\approx 0.86$},
the gap of the system closes at two valleys $\bsl{K}_\pm=(\pi, \pm k_{y0})$ near $X$ along $X-M$ in the BZ, as shown in \figref{fig:bms}(b).
This GC results in a TQPT between the QSH state and the NI state in one Ba-Sb layer, as confirmed by the direct calculation of $Z_2$ index (\figref{fig:bms}(c)) according to expression in \refcite{Fu2006Z2Pumping}.
Since the two GC momenta are invariant under $\TR m_y$, this GC case satisfies the definition of scenario (iii) for $p1m1$.
We further numerically verify the PET jump induced by the GC with the tight-binding model.
The jump of the symmetry-allowed PET components is found at the TQPT around {$\alpha=\alpha_\text{c}$} in \figref{fig:bms}(d), while the components forbidden by the symmetry stay zero.
According to \figref{fig:bms}(d), both the jump and background are of the same order of magnitude, 0.1 e{$\AA^{-1}$} for $\gamma_{yxy,yyx}$ and  0.01 e{$\AA^{-1}$} for $\gamma_{xxx,xyy}$, indicating that the jump is experimentally measurable.
The $Z_2$ topology change and the PET jump can also be analytically verified based on the effective model discussed in {Appendix F}.

{
\section{Discussion}
\label{sec:conclusion}
}
In conclusion, we demonstrate that for all PGs that allow nonvanishing PET, the piezoelectric response has a discontinuous change across any TQPT in 2D TR invariant systems with significant SOC.
Potential material realizations include the HgTe/CdTe quantum well and the layered material BaMnSb$_2$.

The early study on MoS$_2$ has demonstrated that the values of the PET obtained from the effective model might be (though not always) quite close to those from the first principles calculations~\cite{Rostami2018PiE}.
Therefore, although our theory is based on the effective Hamiltonian, the predicted jump of the PET is quite likely to be significant and even the sign change of PET, such as \figref{fig:HgTe}(b) and (d) for the HgTe case, might exist in realistic materials.
The evaluation of the PET from the first principles calculations is left for the future works.

Although we only focus on two realistic material systems in this work, the theory can be directly applied to other material systems.
For example, the calculations for the HgTe/CdTe QW are also applicable to InAs/GaSb QWs, which share the same model~\cite{Knez2011InAsGaSb}.
The QSH effect has also been observed in the monolayer 1T'-WTe$_2$~\cite{Tang2017QSH,Fei2017QSH,Wu2018QSH}, but its inversion symmetry~\cite{Qian2014QSHTMD} forbids the piezoelectric effect.
Therefore, a significant inversion breaking effect from the environment (such as substrate) is required to test our prediction in this system.
While the SOC strength in graphene is small, it has been shown that the bilayer graphene sandwiched by TMDs has enhanced SOC and serves as a platform to observe TQPT~\cite{Island2019BLGSOC,Zaletel2019BLGSOC}, where the PET jump is likely to exist.
The piezoelectric effect has been observed in several 2D material systems~\cite{Zhu2014PET,Wu2014PET,Fei2015PET}, and therefore, the material systems and the experimental technique for the observation of the PET jump are both available.
Since the PET jump is directly related to the TQPT, it further provides a new experimental approach to extract the critical exponents and universality behaviors of the TQPT, which can only be analyzed through transport measurements nowadays.

This work only focuses on 2D TR invariant systems with SOC, and the generalization to systems without SOC, without TR symmetry, or in 3D is left for the future.
Despite the similarity between \eqnref{eq:gamma_bc} and the expression of CN, the generalization to TR-breaking systems with non-zero CNs requires caution, due to the change of the definition of polarization~\cite{Vanderbilt2009PolarizationCI}.
Another interesting question is whether the PET jump exists across the transition between states of different higher-order~\cite{Benalcazar2017HOTI,Schindler2018HOTI,Song2017HOTI,
Langbehn2017SOTITSC} or fragile topology~\cite{Po2018FragileTopo,
Bradlyn2019Fragile}.
We notice that although the dynamical piezoelectric effect may exist in metallic systems~\cite{Varjas2016PET}, its description is different from \eqnref{eq:gamma_bc}. 
It is thus intriguing to ask how the dynamical PET behaves across the transitions between insulating and semimetal phases.

\section{Methods}

%

\subsection{Expression for the PET}
According to \refcite{Vanderbilt2000BPPZ,Vanderbilt2009PolarizationCI}, the expression for the PET of insulators, \eqnref{eq:gamma_bc}, is derived for systems with zero CNs and within the clamped-ion approximation where ions exactly follow the homogeneous deformation and thus cannot contribute to the PET.
Even though the ion contribution might be non-zero in reality, the approximation is still legitimate in our study of PET jump since the ion contribution varies continuously across the GC of electronic bands.

\eqnref{eq:gamma_bc} involves the derivative of the periodic part of the Bloch state $\ket{\varphi_{n,\bsl{k}}}$ with respect to the strain tensor $u_{jk}$.
$\ket{\varphi_{n,\bsl{k}}}$ can always be expressed as $\ket{\varphi_{n,\bsl{k}}}=\sum_{\bsl{G}}f_{n,\bsl{k},\bsl{G}}\ket{\bsl{G}}$ with $\bsl{G}$ the reciprocal lattice vector, and the derivative in fact means $\ket{\partial_{u_{jk}}\varphi_{n,\bsl{k}}}\equiv \sum_{\bsl{G}}(\partial_{u_{jk}}f_{n,\bsl{k},\bsl{G}})\ket{\bsl{G}}$~\cite{Vanderbilt2000BPPZ}.
In this way, the ill-defined $\partial_{u_{ij}}\ket{\bsl{G}}$ is avoided, despite that $\ket{\bsl{G}}$ is not continuous as changing the strain.
If replacing the $\ket{\partial_{u_{jk}}\varphi_{n,\bsl{k}}}$ in \eqnref{eq:gamma_bc} by a momentum derivative $\ket{\partial_{k_j}\varphi_{n,\bsl{k}}}$ with $j$ different from $i$, the PET expression transforms into $-e C \epsilon^{ij}/(2\pi)$, where $\epsilon^{ij}=-\epsilon^{ji}$, $\epsilon^{xy}=1$, and $C$ is the Chern number of the 2D insulator~\cite{TKNN}
\eq{
C=\int \frac{d^2k}{2\pi} \sum_{n} F_{k_x, k_y}^n\ .
}
This reveals the similarity between the PET expression and the expression of the CN.

\subsection{PG $p1$}
For $p1$, no special constraints are imposed on the PET.
There are two GC scenarios for the PG $p1$ with TR symmetry:
\begin{itemize}
\item (i) gap closes at TRIM ($\TR\in\mathcal{G}_0$),
\item (ii) gap closes not at TRIM ($\TR\notin\mathcal{G}_0$).
\end{itemize}
In scenario (ii), $\mathcal{G}_0$ contains no symmetries other than the lattice translation, which we refer to as the trivial $\mathcal{G}_0$.

\subsection{PGs $p1m1,c1m1$, and $p1g1$}
All three PGs, $p1m1,c1m1$, and $p1g1$, are generated by a mirror-related symmetry $\mathcal{U}$ and the lattice translation.
$\mathcal{U}$ is a mirror operation for $p1m1/c1m1$ and a glide operation for $p1g1$.
The difference between $p1m1$ and $c1m1$ lies on the directions of the primitive lattice vectors relative to the mirror line, which is not important for our discussion here.
Without loss of generality, we choose the mirror or glide line to be perpendicular to $x$, labelled as $m_x$ or $g_x$, respectively.
The glide operation is thus denoted as $g_x=\{m_x|0\frac{1}{2}\}$, where $0\frac{1}{2}$ represents the translation by half the primitive lattice vector along $y$.
The $\mathcal{U}$ symmetry in these three PGs requires
\eq{
\label{eq:gamma_mirror}
\gamma_{ijk}=(-1)^i(-1)^j(-1)^k \gamma_{ijk}
}
with $(-1)^x=-1$ and $(-1)^y=1$, resulting that $\gamma_{xxx}=\gamma_{xyy}=\gamma_{yxy}=\gamma_{yyx}=0$ while $\gamma_{xxy}, \gamma_{xyx}, \gamma_{yxx}, \gamma_{yyy}$ are allowed to be nonzero.
For the symmetry analysis here, the PET behaves the same under the glide and mirror operations since $u_{ij}$ is considered in the continuum limit.
Based on $\mathcal{G}_0$, we obtain in total 4 GC scenarios for these three PGs:
\begin{itemize}
\item (i) the GC at TRIM ($\mathcal{G}_0$ contains $\TR$),
\item (ii) $\mathcal{G}_0$ contains $\mathcal{U}$ but not $\TR$,
\item (iii) $\mathcal{G}_0$ contains $\mathcal{U}\TR$ but not $\TR$,
\item (iv) $\mathcal{G}_0$ is trivial.
\end{itemize}
\subsection{PG $p3$}
PG $p3$ is generated by 3-fold rotation $C_3$ and the lattice translation.
Owing to $C_3$, the PET satisfies the following relation
\eq{
\label{eq:gamma_C3}
\gamma_{ijk}= \sum_{i'j'k'}[R(C_3)]_{ii'}[R(C_3)]_{jj'}[R(C_3)]_{kk'}\gamma_{i'j'k'}\ ,
}
where
\eq{
R(C_3)=\left(
\begin{array}{cc}
 -\frac{1}{2} & -\frac{\sqrt{3}}{2} \\
 \frac{\sqrt{3}}{2} & -\frac{1}{2} \\
\end{array}
\right)\ .
}
Solving the above equation gives two independent components $\gamma_{xxx}$ and $\gamma_{yyy}$ as
\eqa{
\label{eq:PET_sym_P3}
&\gamma_{yxy}=\gamma_{yyx}=\gamma_{xyy}=-\gamma_{xxx}\\ &\gamma_{xxy}=\gamma_{xyx}=\gamma_{yxx}=-\gamma_{yyy}\ .
}
Again, we classify the GC for $p3$ according to $\mathcal{G}_0$, resulting in three different scenarios:
\begin{itemize}
\item (i) $\mathcal{G}_0$ contains $\mathcal{T}$,
\item (ii) $\mathcal{G}_0$ contains $C_3$ but not $\mathcal{T}$,
\item (iii) $\mathcal{G}_0$ is trivial.
\end{itemize}
Here we do not have a scenario for $\mathcal{G}_0$ containing $C_3\mathcal{T}$ but no $\mathcal{T}$, since $(C_3\mathcal{T})^3$ is equivalent to $\mathcal{T}$.

\subsection{PGs $p31m$ and $p3m1$}

Both PGs $p31m$ and $p3m1$ are generated by the lattice translation, the three-fold rotation $C_3$, and a mirror symmetry which we choose to be $m_x$ without loss of generality.
The difference between the two PGs lies on the direction of the mirror line relative to the primitive lattice vector: the mirror line is parallel or perpendicular to one primitive lattice vector for $p31m$ or $p3m1$, respectively.
$C_3$ and $m_x$ span the point group $C_{3v}$, which makes the PET satisfy \eqnref{eq:gamma_mirror} and \eqnref{eq:gamma_C3}.
As a result, we have
\eqa{
\label{eq:PET_sym_p3m}
& \gamma_{xxx}=\gamma_{xyy}=\gamma_{yxy}=\gamma_{yyx}=0\\
& \gamma_{xyx}=\gamma_{xxy}=\gamma_{yxx}=-\gamma_{yyy}
}
for the PET, and thus $\gamma_{yyy}$ serves as the only independent symmetry-allowed PET component.
We classify the GC scenarios into 4 types according to $\mathcal{G}_0$:
\begin{itemize}
\item (i) $\mathcal{G}_0$ contains $\mathcal{T}$,
\item (ii) $\mathcal{G}_0$ contains at least one of the three mirror symmetry operations in $C_{3v}$ (again labeled as $\mathcal{U}=m_x$, $C_3 m_x$, or $C_3^2 m_x$) but no $\mathcal{T}$,
\item (iii) $\mathcal{G}_0$ contains $\mathcal{U}\mathcal{T}$ but no $\mathcal{T}$,
\item (iv) $\mathcal{G}_0$ is trivial.
\end{itemize}

\subsection{Valley CN}
In all the valley CN cases (\figref{fig:GC}(d,e,n,o)), the GC points locate at generic positions in the 1BZ.
The valleys can be physically defined as the positions where the Berry curvature diverges as the gap approaches to zero.
The positions of the Berry curvature peaks around the gap closing can be clearly seen in numerical calculations, as long as those peaks are well separated in the momentum space. (See {Appendix D} for more details.)
With the positions of the valleys determined, the valley CN on one side of the GC is not necessarily quantized to integers since the integral of Berry curvature is not over a closed manifold.
However, the change of valley CN across the GC is always integer-valued, since it is equal to the CN of the Hamiltonian given by patching the two low-energy effective models on the two sides of the GC at large momenta, which lives on a closed manifold.
One physical consequence of the quantized change of valley CN is the gapless domain-wall mode~\cite{Martin2010VCN}, which can be experimentally tested with transport or optical measurements~\cite{Zhu2018ValleyGraphene}.
We verify the quantized change of valley CN and demonstrate the corrsponding gapless domain-wall mode with a tight-binding model in {Appendix D}.

The above argument relies on the constraint that the valleys are well separated in 1BZ, preventing the two states from being adiabatically connected.
Without the contraint of well-defined valleys, the valleys are allowed to be merged, and two phases with different valley CNs might be adiabatically connected.
Therefore, we refer to the topology characterized by valley CN as locally stable~\cite{Fang2015TCI}, though globally unstable.
Nevertheless, we restrict all valleys to be well-defined in our discussion and refer to the corresponding gap closing case as a TQPT.

%

\section{Acknowledgement}
We are thankful for the helpful discussion with B. Andrei Bernevig, Xi Dai, F. Duncan M. Haldane, Shao-Kai Jian, Biao Lian, Xin Liu, Laurens W. Molenkamp, Zhiqiang Mao, Xiao-Qi Sun, David Vanderbilt, Jing Wang, Binghai Yan, Junyi Zhang, and Michael Zaletel.
We acknowledge the support of the Office of Naval Research (Grant No. N00014-18-1-2793), the U.S. Department of Energy (Grant No.~DESC0019064) and Kaufman New Initiative research grant KA2018-98553 of the Pittsburgh Foundation.

\appendix

\section{Derivation of the PET}
\label{app:PET_linear_response}
In this section, we derive \eqgammabc in the main text via linear response theory from \eqPETPro in the main text, which is equivalent to the derivation in \refcite{Vanderbilt2000BPPZ}.
The derivation is done with the natural unit $c=\hbar=1$ and the metric $(-,+,+)$.

To apply the linear response theory, we start from an action $S$ that includes the electronic effective model and the leading order effect of the infinitesimal strain.
Since the current is present in \eqPETPro, we should include the $U(1)$ gauge field that accounts for the electromagnetic field.
With the $U(1)$ gauge field, the action reads
\eq{
S=\int \frac{d^3 k }{(2\pi)^3}\psi^\dagger_{k} G^{-1}_0(k)\psi_{k}+\int \frac{d^3 k }{(2\pi)^3}\int \frac{d^3 q }{(2\pi)^3}\left[ \psi^\dagger_{k+q/2} \frac{\partial G^{-1}_0(k)}{\partial k^\mu}\psi_{k-q/2} e A^\mu(q)-\psi^\dagger_{k+q/2} M_{ij} \psi_{k-q/2} u_{ij}(q) \right]\ ,
}
where $k^\mu=(\omega,\bsl{k})_\mu$, $A^\mu$ and $u_{ij}$ and $\psi$ follow the same Fourier transformation rule, $G_0(k)=[\omega-h_0(\bsl{k})(1-\ii \epsilon)]^{-1}$ is the time-ordered Green function without the electron-strain coupling, the chemical potential is chosen to be the zero energy, and $M_{ij}$ is the matrix coupled to the strain tensor $u_{ij}$.
To the leading order, the linear response is given by the following effective action
\eq{
\label{eq:Seff_PET}
S_{eff}=\int d^3 x\ e \partial_\nu A_\mu u_{ij} f^{ij,\mu\nu}\ ,
}
where
\eqa{
f^{ij,\mu\nu}=&-\frac{1}{2}\int \frac{d^3 k}{(2\pi)^3}\{ \Tr[G_0\frac{\partial G^{-1}_0}{\partial k_\mu} G_0 \frac{\partial G_0^{-1}}{\partial k_\nu}  G_0 M_{ij}]\\
&-(\mu\leftrightarrow\nu)\}\ ,
}
and the absence of the Chern-Simons term $A dA$ is due to the $\TR$ symmetry.

With \eqnref{eq:Seff_PET} and \eqPETPro, we can use the condition that $u_{ij}$ is uniform to derive the expression of the PET, resulting in
\eq{
\label{eq:gamma_GM}
\gamma_{ijk}=-e f^{jk,i0}\ .
}
To further derive \eqgammabc, we define $h(\bsl{k},u_{ij})= h_{0}(\bsl{k})+ u_{ij} M_{ij}$ and $G(k,u_{ij})=[\omega- h(\bsl{k},u_{ij})(1-\ii \epsilon)]^{-1}$ as the Hamiltonian and Green function with the electron-strain coupling, respectively.
Using $\partial_{k_\mu} G^{-1}=\partial_{k_\mu} G^{-1}_0$ and $\partial_{u_{ij}} G^{-1}=-M_{ij}$, we can revise \eqnref{eq:gamma_GM} to
\eqa{
\gamma_{ijk}=&\frac{e }{2}\int \frac{d^3 k}{(2\pi)^3}\{ \Tr[G\frac{\partial G^{-1}}{\partial k_i} G \frac{\partial G^{-1}}{\partial \omega}  G \frac{\partial G^{-1}}{\partial u_{jk}}]\\
&-(k_i\leftrightarrow \omega)\}|_{u_{ij}\rightarrow 0}\ .
}
Define $X^\mu=(\omega,k_i, u_{jk})$ and then the above equation can be further transformed to
\eqa{
&\gamma_{ijk}=\\
&-\left.\frac{e }{3!}\int \frac{d^2 k d\omega}{(2\pi)^3}\epsilon^{\mu\nu\rho}\Tr[G\frac{\partial G^{-1}}{\partial X^\mu} G \frac{\partial G^{-1}}{\partial X^\nu}  G \frac{\partial G^{-1}}{\partial X^\rho}]\right|_{u_{ij}\rightarrow 0}\ ,
}
where $\epsilon^{\mu\nu\rho}$ is the Levi-Civita symbol.
Integrating out $\omega$ in the above equation with the Wick rotation gives \eqgammabc.
Although the derivation here is done for $\hbar=c=1$, all the expressions of $\gamma_{ijk}$ and the resultant \eqgammabc stay the same after converting to the SI unit as they carry the right unit for the PET in 2+1D.

Finally, we would like to discuss the effect of the identity term of $h_0$ in \eqnref{eq:Seff_PET} when $h_0$ is a two band model.
In general, the Hamiltonian can always be split into the identity part and the traceless part as $h_0(\bsl{k})=m_0(\bsl{k})\mathds{1}+h_0^{\text{traceless}}(\bsl{k})$.
The eigenvalues of $h_0(\bsl{k})$ then read $m_0(\bsl{k})\pm \varepsilon(\bsl{k})$, where $\pm \varepsilon(\bsl{k})$ are two eigenvalues of $h_0^{\text{traceless}}(\bsl{k})$ with $\varepsilon(\bsl{k})>0$ chosen without loss of generality.
As the model is gapped and the Fermi energy $(E=0)$ is chosen to lie inside the gap, we have $\varepsilon(\bsl{k})>|m_0(\bsl{k})|\geq 0$.
Since the poles of $G_0$ are at $\omega= [m_0(\bsl{k})\pm \varepsilon(\bsl{k})](1-\ii \epsilon)$, integrating $\omega$ along $(-\infty,\infty)$ in $f^{ij,\mu\nu}$ of \eqnref{eq:Seff_PET} gives the same result as integrating $\omega$ along $(-\infty+ m_0(\bsl{k})(1-\ii \epsilon),\infty+m_0(\bsl{k})(1-\ii \epsilon))$ owing to the absence of poles in between the two paths.
As a result, we can directly neglect the identity term of a two-band insulating $h_0$ in $f^{ij,\mu\nu}$ of \eqnref{eq:Seff_PET}.

\section{Details on PET for Each PG}

The discussion on the electronic effective model and FTP of the gap closing between two non-degenerate states has some overlap with \refcite{Yang20172DGC}.
However, the topological property and PET jump of the gap closing between two insulating states have not been discussed in \refcite{Yang20172DGC}.

\subsection{PG $p1$}
In the main text, the effective Hamiltonian for scenario (ii) of $p1$ is derived in a non-Cartesian coordinate system, which is not convenient for the generalization to other PGs with more crystalline symmetries.
Thus, we re-derive the effective Hamiltonian in the Cartesian coordinate system, as given by (see \appref{app:p1})
\eqa{
\label{eq:h_+-_p1}
 h_{\pm}(\bsl{q},u)= & E_0(\pm\bsl{q})\sigma_0+(v_x q_x+v_0 q_y)\sigma_x+v_y q_y\sigma_y\\
&  \pm m \sigma_z + \xi_{0,ij}\sigma_0 u_{ij}\pm \xi_{a',ij}\sigma_{a'} u_{ij}\ .
}
Here we only perform the unitary transformation on the bases of the Hamiltonian and do not rotate the momentum or the coordinate system.
Correspondingly, the PET jump across the direct TQPT at $m=0$ can be derived as
\eqa{
\label{eq:PET_jump_p1}
&\Delta\gamma_{xij}=-e\frac{\Delta N_+}{\pi} \frac{\xi_{y,ij}}{v_y} \\
&\Delta\gamma_{yij}=e\frac{\Delta N_+}{\pi} \left(\frac{\xi_{x,ij}}{v_x}-\frac{v_0}{v_x} \frac{\xi_{y,ij}}{v_y}\right)\ .
}
\eqnref{eq:h_+-_p1}-\eqref{eq:PET_jump_p1} resemble the conclusion for $p1$ in the Results and are useful for the discussion of the other 6 PGs with non-vanishing PET.

We would like to discuss more about the GC and PET for $p1$.
In the first scenario, all TRIM have no essential differences and the gap closing always happens between two Kramers pairs unless more parameters are finely tuned.
Therefore, there is no need to further classify this scenario into finer cases, and the codimension for the gap closing is 5, indicating that this scenario cannot be direct TQPT~\cite{Murakami2007QSH}.
According to the main text, no finer classification is needed for the second scenario either, the codimension of the gap closing scenario is 1, and it is indeed a direct TQPT that changes the $Z_2$ index and leads to the PET jump.

\subsection{PGs $p1m1$, $c1m1$ and $p1g1$}

\label{sec:pm}

\begin{figure}[t]
    \centering
    \includegraphics[width=0.5\columnwidth]{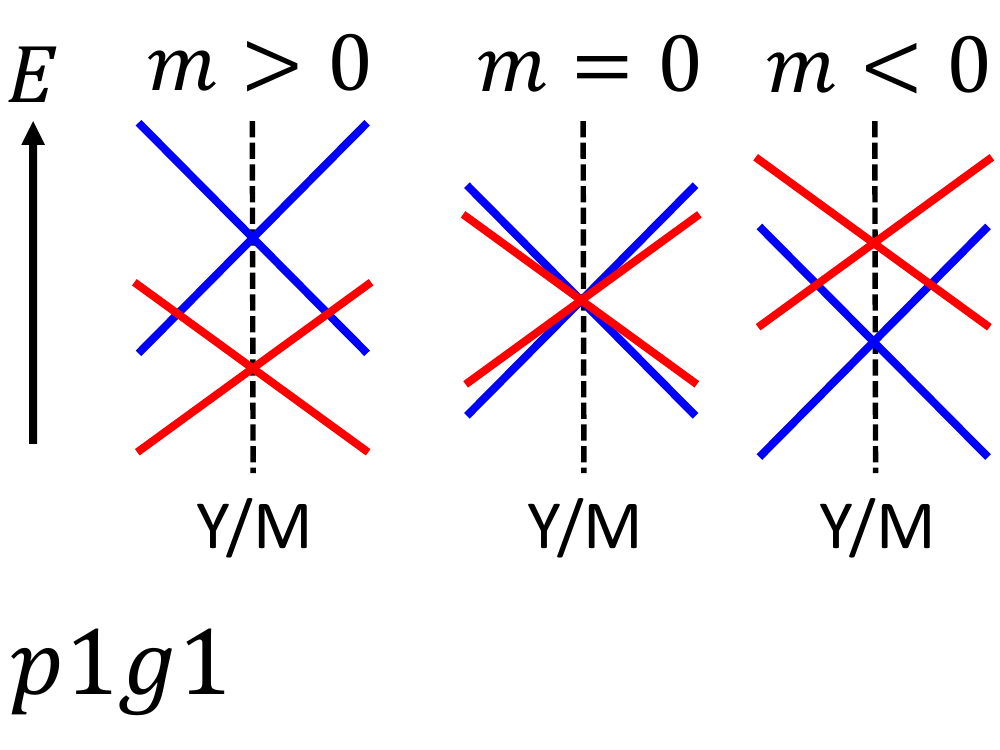}
    \caption{The figure shows the gap closing at $Y$ or $M$ in scenario (i) for $p1g1$.
    The lines indicate the band dispersion along $k_y$, and those in the same (different) colors have the same (opposite) $g_x$ eigenvalues.
    $m$ labels the gap at $Y$ or $M$, and when the gap at $Y$ or $M$ is open ($m\neq 0$), the system is still in a gapless phase.
    }
    \label{fig:SM_pg}
\end{figure}

In this part, we study three PGs, $p1m1,c1m1$, and $p1g1$, all of which are generated by a mirror-related symmetry $\mathcal{U}$ and the lattice translation.
$\mathcal{U}$ is a mirror operation for $p1m1/c1m1$ and a glide operation for $p1g1$.
The difference between $p1m1$ and $c1m1$ lies on the directions of the primitive lattice vectors relative to the mirror line, which is not important for our discussion here.
Without loss of generality, we choose the mirror or glide line to be perpendicular to $x$, labelled as $m_x$ or $g_x$, respectively.
The glide operation is thus denoted as $g_x=\{m_x|0\frac{1}{2}\}$, where ``$0\frac{1}{2}$" represents the translation by half the primitive lattice vector along $y$.
The $\mathcal{U}$ symmetry in these three PGs requires $\gamma_{xxx}=\gamma_{xyy}=\gamma_{yxy}=\gamma_{yyx}=0$, whereas the PET components $\gamma_{xxy}, \gamma_{xyx}, \gamma_{yxx}, \gamma_{yyy}$ are allowed to be nonzero.
For the symmetry analysis here, the PET behaves the same under the glide and mirror operations since $u_{ij}$ is considered in the continuum limit.

In order to classify the gap closing scenarios, we define the group $\mathcal{G}_0$ for a gap closing momentum $\bsl{k}_0$ such that $\mathcal{G}_0$ contains all symmetry operations that leave $\bsl{k}_0$ invariant.
Since $\mathcal{G}_0$ can include the TR-related operation, it can be larger than the little group of $\bsl{k}_0$.
Based on $\mathcal{G}_0$, we obtain in total 4 gap closing scenarios for these three PGs: (i) the gap closing at TRIM ($\mathcal{G}_0$ contains $\TR$), (ii) $\mathcal{G}_0$ contains $\mathcal{U}$ but not $\TR$, (iii) $\mathcal{G}_0$ contains $\mathcal{U}\TR$ but not $\TR$, (iv) $\mathcal{G}_0$ contains no symmetries other than the lattice translation, which we refer to as the trivial $\mathcal{G}_0$.
As summarized in \tabmainresults in the main text, the TQPT exists in scenario (iii) and (iv), which can lead to the jump of symmetry-allowed PET components.

\subsubsection{Scenario (i): TRIM}

\label{sec:pm_i}

In scenario (i), the gap closing requires 3 (5) fine-tuning parameters (FTPs) for $p1m1$ and $c1m1$ if $m_x$ is (is not) in the $\mathcal{G}_0$. (See \appref{app:pm}.)
For $p1g1$, the TRIM ($\Gamma$, $X$, $Y$ and $M$) are split into two classes according to the value of $g_x^2$: $\Gamma,X$ with $g_x^2=-1$ and $Y,M$ with $g_x^2=1$.
The gap closing at $\Gamma,X$ needs 3 FTPs since $g_x$ behaves the same as $m_x$, while the gap closing at $Y,M$ needs only 1 FTP if it happens between two Kramers pairs with opposite $g_x$ eigenvalues.
However, such gap closing at $Y,M$ is in between two $g_x$-protected gapless phases with codimension 0, where the bands with opposite $g_x$ eigenvalues cross with each other at momenta other than $Y,M$ as shown in \figref{fig:SM_pg}.
Therefore, there is no direct TQPT between two gapped phases in scenario (i).

\subsubsection{Scenario (ii): $\mathcal{U}\in \mathcal{G}_0$ but $\TR\notin \mathcal{G}_0$}

\label{sec:pm_ii}

The same situation occurs for scenario (ii).
In scenario (ii), the gap closes at two different momenta $\pm \bsl{k}_0$ that are invariant under the $\mathcal{U}$ operation, meaning that the bases at $\pm \bsl{k}_0$ can have definite $\mathcal{U}$ eigenvalues.
The gap closing between the two bases with the same $\mathcal{U}$ eigenvalues requires 2 FTPs, as discussed in \appref{app:pm}.
When the gap closes between two bands with opposite $\mathcal{U}$ eigenvalues, the system always enters a stable $\mathcal{U}$-protected gapless phase with 0 codimension.
(This case is not the same as the scenario (i) since only one side is guaranteed to be gapless.)
Thus, the gap closing cases cannot be direct TQPTs.

\subsubsection{Scenario (iii): $\mathcal{U}\TR\in \mathcal{G}_0$ but $\TR\notin \mathcal{G}_0$}

\label{sec:pm_iii}

In scenario (iii), the gap closing occurs at two different momenta $\pm\bsl{k}_0$ that are invariant under $\mathcal{U}\mathcal{T}$, as shown by the orange dashed lines in \figGC(b) and (c) in the main text.

For $p1m1$ and $c1m1$ ($\mathcal{U}=m_x$), $(m_x\mathcal{T})^2=1$ suggests that we can have $m_x\mathcal{T}\dot{=} \cc$ at $\bsl{k}_0$ by choosing the appropriate bases and the band touching point at $\bsl{k}_0$ should typically occur between two non-degenerate bands.
We further take $\mathcal{T}\dot{=}\ii\sigma_y \cc$ by choosing the appropriate bases at $-\bsl{k}_0$, and thus the two-band effective models $h_\pm(\bsl{q},u)$ at $\pm\bsl{k}_0$ can be given by \eqnref{eq:h_+-_p1} with extra constraints
\eq{
\label{eq:para_pm_iii}
v_0=\xi_{a_1,xy}=\xi_{a_1,yx}=\xi_{y,xx}=\xi_{y,yy}=0
}
for $a_1=0,x,z$.
As a result, only 1 FTP $m$ is needed for the gap closing ($m=0$), and only one single Dirac cone exists in half 1BZ at the transition, leading to the change of the $Z_2$ index.
Based on \eqnref{eq:PET_jump_p1}, the jump of symmetry-allowed PET components across this TQPT can be derived as
\eqa{
\label{eq:PET_jump_pm}
&\Delta\gamma_{xxy}=\Delta\gamma_{xyx}=-e\frac{\Delta N_+}{\pi} \frac{\xi_{y,xy}}{v_y} \\
&\Delta\gamma_{yxx}=e\frac{\Delta N_+}{\pi} \frac{\xi_{x,xx}}{v_x}\\
&\Delta\gamma_{yyy}=e\frac{\Delta N_+}{\pi} \frac{\xi_{x,yy}}{v_x}\ .
}

For $p1g1$ with $\mathcal{U}=g_x$, since $(g_x\mathcal{T})^2=1$ at $(k_x,0)$ and $(g_x\mathcal{T})^2=-1$ at $(k_x,\pm\pi)$, we have two different gap closing cases.
When the gap closes at $(\pm k_{0,x},0)$, the algebra relation involving $g_x\mathcal{T}$ is the same as $m_x\mathcal{T}$, \eg $(g_x\mathcal{T})^2=(m_x\mathcal{T})^2=1$, and thus the effective Hamiltonian can be chosen to be the same as that for $p1m1$ and $c1m1$, leading to 1 FTP, $Z_2$ index change, and the same form of PET jump.
On the contrast, due to $(g_x\mathcal{T})^2=-1$ at $(\pm k_{0,x},\pm\pi)$, the gap closing needs 4 FTPs and thus no TQPT can occur in this case. (See \appref{app:pm}.)

\subsubsection{Scenario (iv): trivial $\mathcal{G}_0$}

\label{sec:pm_iv}

In scenario (iv), the gap should close simultaneously at four momenta $\bsl{k}_0$, $\bsl{k}_1=-\bsl{k}_0$, $\bsl{k}_2=\mathcal{U} \bsl{k}_0$, and $\bsl{k}_3=-\mathcal{U} \bsl{k}_0$, as depicted in \figGC(d) and (e) in the main text.
The gap closing at $\bsl{k}_0$ can be described by the Hamiltonian $h_{+}(\bsl{q},u)$ in \eqnref{eq:h_+-_p1}, and the Hamiltonian at $\bsl{k}_1$, $\bsl{k}_2$, and $\bsl{k}_3$ can be given by
$\mathcal{T}h_{+}(-\bsl{q},u)\mathcal{T}^{\dag}$, $\mathcal{U}h_{+}(\mathcal{U}^{-1}\bsl{q},\mathcal{U}^{-1}u (\mathcal{U}^{-1})^T)\mathcal{U}^{\dag}$, and $\mathcal{UT}h_{+}(-\mathcal{U}^{-1}\bsl{q},\mathcal{U}^{-1}u (\mathcal{U}^{-1})^T)(\mathcal{UT})^{\dag}$, respectively.
Therefore, the gap closing can be achieved by tuning 1 FTP, \ie $m$ in $h_{+}(\bsl{q},u)$, in this scenario.

There is no change of $Z_2$ index for this scenario, since two Dirac cones exist in half 1BZ when the gap closes and the CN of contracted half 1BZ can only change by an even number.
Nevertheless, scenario (iv) can still be ``topological" in the context of valley Chern number (VCN) as elaborated in the following.
Due to the Dirac Hamiltonian form shown in \eqnref{eq:h_+-_p1}, the Berry curvature is peaked at each valley $\bsl{k}_{0,1,2,3}$ for a small $m$ and can be captured by the electronic part of the corresponding effective Hamiltonian.
Then, we can integrate the Berry curvature given by the effective model and get the VCN~ \cite{Zhang2013VCN,Rostami2018PiE} for each valley as $N_{\bsl{k}_i}=-\eta_i \sgn{v_x v_y}\sgn{m}/2$ with $i=0,1,2,3$.
The values of $\eta_i$ at different valleys are related by the TR and $\mathcal{U}$ symmetries, both of which flip the sign of the Berry curvature.
Thus, we have $\eta_0=\eta_3=1$ and $\eta_1=\eta_2=-1$.
It should be pointed out that the Berry curvature integral is not over the entire 1BZ and the VCN at each valley thus does not need to be an integer.
Nevertheless, {\it the change of VCN} across the gap closing is defined on a closed manifold and must be an integer number, given by $\Delta N_{\bsl{k}_i}=-\eta_i \sgn{v_x v_y}$ as varying $m$ from $0^-$ to $0^+$.
For the convenience of further discussion, we can define the VCN of the whole system~\cite{Rostami2018PiE} as $N_{\text{val}}=\sum_i\eta_i N_{\bsl{k}_i}=-2 \sgn{v_x v_y} \sgn{m}$, and the change of the VCN becomes $\Delta N_{\text{val}}= -4 \sgn{v_x v_y}=4 \Delta N_+$ with the factor $4$ for the four valleys.
Therefore, if we restrict all the valleys to be far apart in the momentum space, the change of the VCN is a well-defined topological invariant and this gap closing scenario is a TQPT.

In principle, tuning parameters may merge different valleys at some high symmetry momentum, \eg the valleys at $\bsl{k}_0$ and $\mathcal{U} \bsl{k}_0$ merged at the mirror or glide line.
Therefore, without the constraint of well-defined valleys, two phases with different VCNs can share the same band topology and thus can be adiabatically connected.
It means the topology characterized by VCN is ``locally stable"~\cite{Fang2015TCI}, though globally unstable.
Nevertheless, we restrict all valleys to be well-defined in our discussion and refer to the gap closing scenario as a TQPT.

Next we study the change of the PET components at this TQPT, which can be split into two parts: $\Delta \gamma^{(0)}$ originating from $\pm \bsl{k}_0$ and  $\Delta \gamma^{(1)}$ given by $\pm \mathcal{U}\bsl{k}_0$.
$\Delta \gamma^{(0)}$ equals to \eqnref{eq:PET_jump_p1} since the effective models at $\pm \bsl{k}_0$ are the same as \eqnref{eq:h_+-_p1}.
Owing to the mirror or glide symmetry, $\Delta \gamma^{(1)}$ is related to $\Delta \gamma^{(0)}$ as $\Delta \gamma^{(1)}_{ijk}=(\mathcal{U})_{ii'}(\mathcal{U})_{jj'}(\mathcal{U})_{kk'}\gamma^{(0)}_{i'j'k'}$.
As a result, we obtain the non-zero jump of symmetry-allowed PET components $\Delta \gamma_{ijk}=\Delta \gamma^{(0)}_{ijk}+\Delta \gamma^{(1)}_{ijk}$ as
\eqa{
\label{eq:PET_pm_iv}
&\Delta\gamma_{xxy}=\Delta\gamma_{xyx}=-e\frac{ \Delta N_{\text{val}}}{2\pi v_y} \xi_{yxy}\\
&\Delta\gamma_{yxx}=-e\frac{\Delta N_{\text{val}}(-v_y \xi_{xxx}+v_0 \xi_{yxx})}{2\pi v_x v_y}\\
&\Delta\gamma_{yyy}=-e\frac{\Delta N_{\text{val}}(-v_y \xi_{xyy}+v_0 \xi_{yyy})}{2\pi v_x v_y}\ .
}

\subsection{PG $p3$ }

\begin{figure}[t]
    \centering
    \includegraphics[width=0.6\columnwidth]{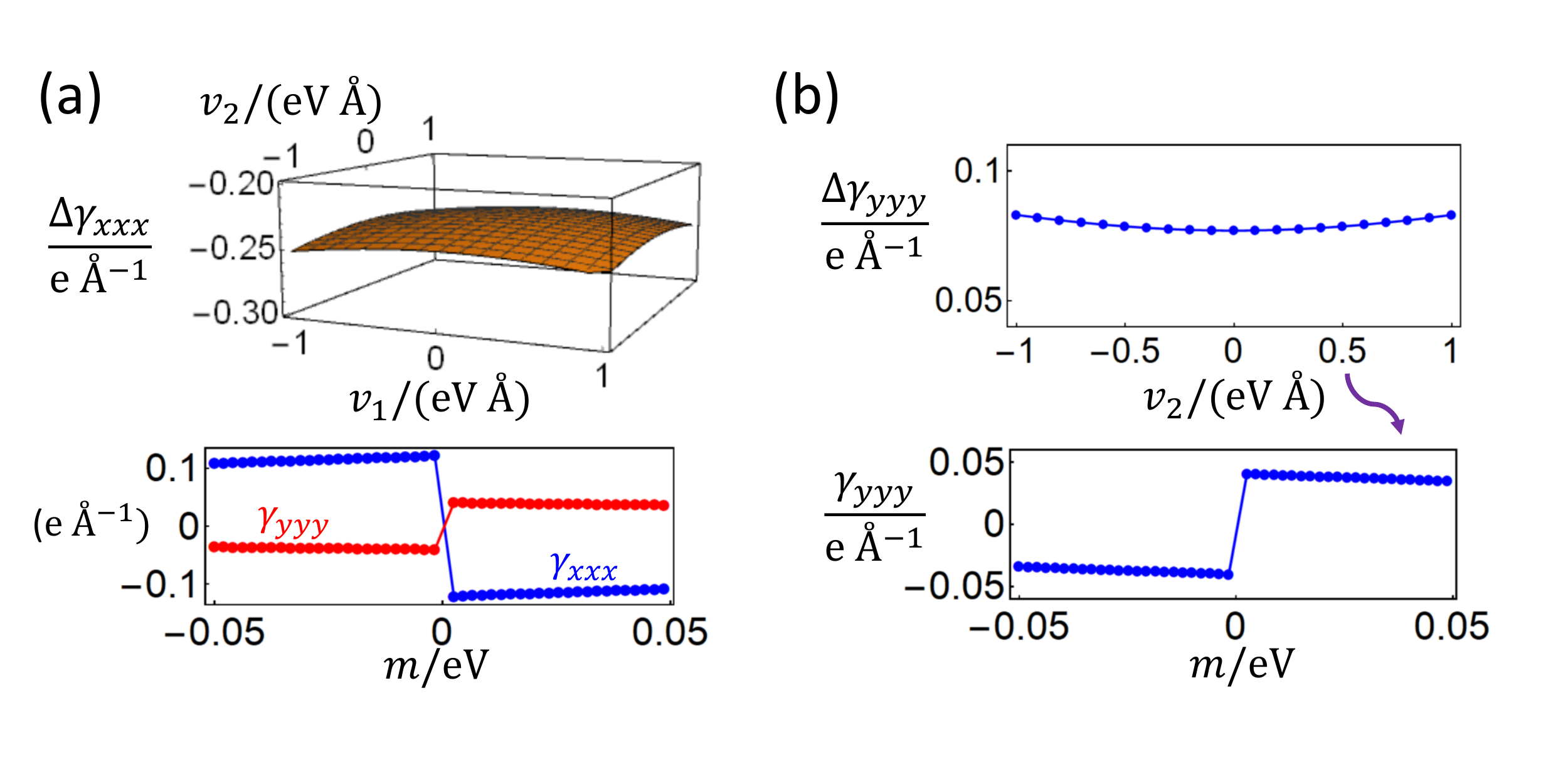}
    \caption{    The upper panels of (\textbf{a}) and (\textbf{b}) numerically show the PET jump induced by the gap closing at $\Gamma$ point as the function of $v_{1,2}$ and $v_2$, respectively.
    The lower panels of (a) and (b) plot the PET component as a function of $m$ for $(v_1,v_2)=(0.5,0.5)$eV$\AA$ and $v_2=0.5$eV$\AA$, respectively.
     }
    \label{fig:SM_Gamma}
\end{figure}

\label{sec:p3}

PG $p3$ is generated by 3-fold rotation $C_3$ and the lattice translation.
Owing to $C_3$, the PET only has two independent components $\gamma_{xxx}$ and $\gamma_{yyy}$ as
\eqa{
\label{eq:PET_sym_P3}
&\gamma_{yxy}=\gamma_{yyx}=\gamma_{xyy}=-\gamma_{xxx}\\ &\gamma_{xxy}=\gamma_{xyx}=\gamma_{yxx}=-\gamma_{yyy}\ .
}
Again, we classify the gap closing for $p3$ according to $\mathcal{G}_0$, resulting in three different scenarios: (i) $\mathcal{G}_0$ contains $\mathcal{T}$, (ii) $\mathcal{G}_0$ contains $C_3$ but not $\mathcal{T}$, and (iii) $\mathcal{G}_0$ is trivial.
Here we do not have a scenario for $\mathcal{G}_0$ containing $C_3\mathcal{T}$ but no $\mathcal{T}$, since $(C_3\mathcal{T})^3$ is equivalent to $\mathcal{T}$.
As summarized in \tabmainresults in the main text and elaborated in the following, in any of the above scenarios, there are gap closing cases between gapped states that need only 1 FTP, change the $Z_2$ index, and lead to the discontinuous change of symmetry-allowed PET components.

\subsubsection{Scenario (i):TRIM}

\label{sec:p3_i}

There are 4 TRIM in scenario (i), namely three $M$ points related by $C_3$ and one $\Gamma$ point, as labeled in \figGC(f) in the main text.
$\mathcal{G}_0$ of each individual $M$ point only contains $\mathcal{T}$ and the lattice translation, and thus the gap closing at $M$ needs 5 FTPs, same as the gap closing at TRIM for $p1$.

When the gap closes at $\Gamma$ point as shown in \figGC(f) in the main text, $\mathcal{G}_0$ also contains $C_3$ with $C_3^3=-1$.
Due to $[C_3,\mathcal{T}]=0$, the Kramers pairs can be classified into two types according to the $C_3$ eigenvalues: one with $(\ee^{- \ii \pi/3},\ee^{ \ii \pi/3})$ and the other with $(-1,-1)$.
The gap closing between the Kramers pairs of the same type requires more than 1 FTPs, 3 for $(\ee^{- \ii \pi/3},\ee^{ \ii \pi/3})$ type and 5 for $(-1,-1)$ type, as discussed in \appref{app:p3}.

The gap closing with 1 FTP happens between the TR pairs of different types, for which the minimal four-band effective Hamiltonian in the bases $(\ee^{- \ii \pi/3},\ee^{ \ii \pi/3},-1,-1)$ reads
\eq{
\label{eq:hp3_i}
h_{p3}(\bsl{k},u)= h_{p3,0}(\bsl{k})+h_{p3,1}(u)\ ,
}
where $h_{p3,0}$ is the electron part
\eqa{
&h_{p3,0}(\bsl{k})= E_0 \tau_0\sigma_0+m\tau_z\sigma_0+(v_1 k_x + v_2 k_y)(\frac{\tau_0+\tau_z}{2})\sigma_x\\
&+(v_1 k_y - v_2 k_x)(\frac{\tau_0+\tau_z}{2})\sigma_y+(v_3 k_x+v_4 k_y) \tau_x\sigma_z\\
&+(-v_3 k_y+v_4 k_x)\tau_y\sigma_0+(v_5 k_x + v_6 k_y)\tau_x\sigma_x\\
&+(-v_5 k_y+ v_6 k_x)\tau_x\sigma_y\ ,
}
and $h_{p3,1}$ describes the electron-strain coupling
\eqa{
&h_{p3,1}(u)=  (u_{xx}+u_{yy})(\xi_1 \tau_0\sigma_0+\xi_2\tau_z\sigma_0)\\
&+(-u_{xx}+u_{yy})(\xi_3\tau_y\sigma_z+\xi_5\tau_y\sigma_x-\xi_4\tau_x\sigma_0+\xi_6 \tau_y\sigma_{y})\\
&+(u_{xy}+u_{yx})(\xi_4\tau_y\sigma_z+\xi_6\tau_y\sigma_x+\xi_3\tau_x\sigma_0-\xi_5 \tau_y\sigma_{y})\ .
}
$\tau$'s and $\sigma$'s are Pauli matrices that label two different Kramers pairs and two components of each Kramers pair, respectively, $m$ is the gap closing tuning parameter, and the bases are chosen such that $\mathcal{T}\dot{=}-\ii\tau_0\sigma_y\mathcal{K}$.

This gap closing is certainly a TQPT since it changes the numbers of IRs of the occupied bands, meaning that the two gapped states separated by this gap closing cannot be adiabatically connected.
When $v_1=v_2=0$, we can define an effective inversion symmetry $\widetilde{P}=\tau_z\sigma_0$ for the electron part of \eqnref{eq:hp3_i}, $\widetilde{P} h_{p3,0}(-\bsl{k}) \widetilde{P}^\dagger=h_{p3,0}(\bsl{k})$, and thus the gap closing of $h_{p3,0}(\bsl{k})$ with $v_1=v_2=0$ changes the $Z_2$ index according to the Fu-Kane criteria~\cite{Kane2007FuKaneInvesion} since the parity of the occupied band changes.
The existence of non-zero $v_1,v_2$ terms that break $\widetilde{P}$ cannot influence the $Z_2$ topology change, since (i) the $Z_2$ topology does not rely on the effective inversion symmetry, and (ii) additional gap closing away from $\Gamma$ is forbidden at $m=0$ as long as the $v_{1,2}$ terms are restored adiabatically.
%
%
Therefore, within a certain range of $v_{1,2}$, the codimension-1 gap closing at $m=0$ is a direct TQPT that changes the $Z_2$ index.

\begin{figure}[t]
    \centering
    \includegraphics[width=0.6\columnwidth]{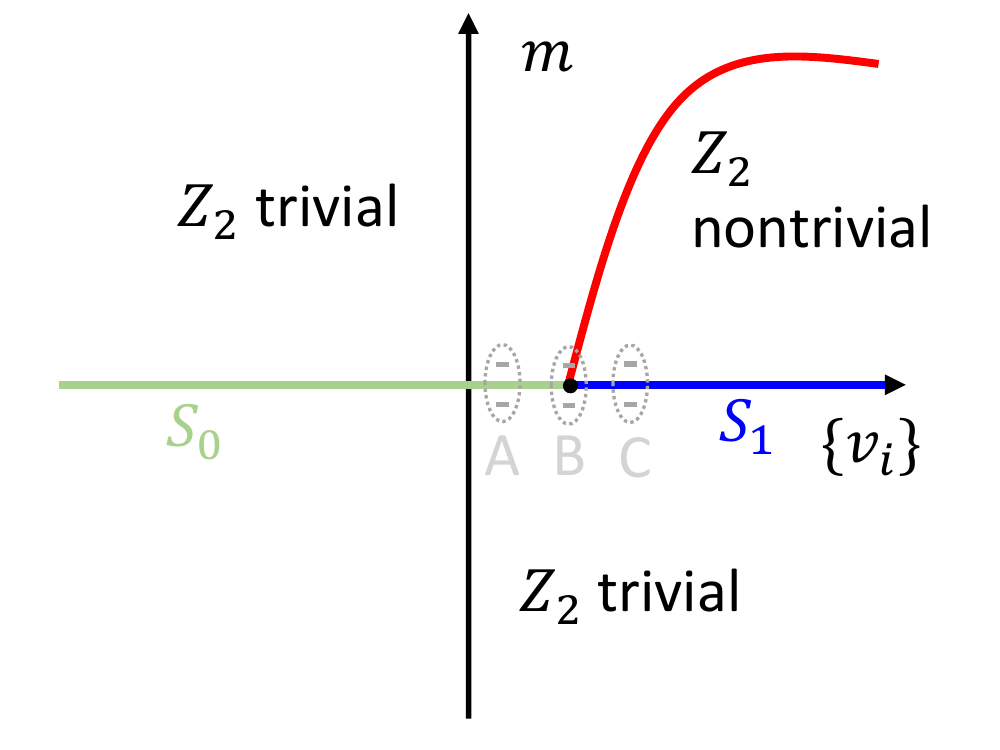}
    \caption{
    This figure shows the typical phase diagram around $(m=0,v_i\in \partial S_0)$ (the black dot), within the assumption that both $S_0$ and $S_1$ are codimension-0 subspaces of the $v_i$ parameter space.
    The green line is $(m=0,v_i\in S_0)$ which does not change $Z_2$ index, the blue line is $(m=0,v_i\in S_1)$ that changes $Z_2$ index, and the system closes the gap at six generic k points on the red line.
    Without loss of generality, we choose the red line to touch the $m=0$ line from the positive $m$ side.
    A,B, and C label three different Hamiltonians given by patching the two effective models with $m=\pm \epsilon$ at different values of $v_i$.
    $A$ is $Z_2$ trivial, $C$ is $Z_2$ non-trivial, and $B$ closes the gap at the six generic $k$ points for in $m=\epsilon$ part.
     }
    \label{fig:SM_p3_Z2}
\end{figure}

%
The remaining question is if the codimension-1 gap closing at $m=0$ is always a $Z_2$ transition.
To answer this question, note that we can always assume the transition at $m=0$ is $Z_2$ for a parameter region $S_1$ of $v_i$'s in $h_{p3,0}$ and non-$Z_2$ for the other parameter region $S_0$ of $v_i$'s.
Since the same form of the Hamiltonian can not correspond to $Z_2$ and non-$Z_2$ transitions simultaneously, the intersection of $S_0$ and $S_1$ is empty.
Now we suppose both $S_0$ and $S_1$ are codimension-0 subspaces of the $v_i$ parameter space (not the whole parameter space since only $v_i$'s are included while $m$ is excluded).
Then, the boundary of $S_0$, labeled as $\partial S_0$, is a codimension-1 subspace of $v_i$ parameter space, and the special transition at $(m=0, v_i \in \partial S_0)$ is a codimension-2 transition, as shown in \figref{fig:SM_p3_Z2}.

Patching the Hamiltonian with $m=\pm \epsilon$ with $\epsilon$ positive and infinitesimal gives a Hamiltonian that lives on a closed manifold.
This Hamiltonian has $Z_2$ trivial and nontrivial ground states when $v_i$'s are in $S_0$ and $S_1$, respectively, as shown in \figref{fig:SM_p3_Z2}.
As $v_i$'s change from $S_0$ to $S_1$ passing through $\partial S_0$, the patched Hamiltonian must experience a gap closing at generic k points that changes $Z_2$ (\figref{fig:SM_p3_Z2}), since there is always an energy gap at $\Gamma$ for $m\neq 0$.
%
As discussed with more detail in the following (see scenario (iii)), the gap closing at generic k points surely changes the $Z_2$ index, is codimension-1, and simultaneously happens at six momenta.
The gap closing can only happen either for $m=\epsilon$ part or for $m=-\epsilon$ part of the patched Hamiltonian but not both, since if the gap closes twice, the $Z_2$ index would be changed back.
%
%
It means, the codimension-1 hypersurface for the gap closing at generic k (red line in \figref{fig:SM_p3_Z2}) touches the codimension-1 hypersurface for gap closing at $\Gamma$ ($m=0$ line in \figref{fig:SM_p3_Z2}) just from one side of $m$ but not passing through.
As mentioned above, the touching part at $(m=0, v_i\in \partial S_0)$ is a codimension-2 transition, owing to the assumption that both $S_0$ and $S_1$ are codimension-0 subspaces of $v_i$ parameter space.

%
At the touching, we must have the six generic gap closing points merging at $\Gamma$.
Otherwise, we should expect the red line in \figref{fig:SM_p3_Z2} to pass through the $m=0$ line instead of stopping, since the gap closing process is local in the momentum space and different gap closing cases cannot influence each other if they happen at the different momenta.
%
However, the merging process cannot be codimension-2 since moving a generic gap closing point to a specific momentum while keeping the gap closed requires at least 3 FTPs (two to move the momentum and one to close gap).
Therefore,  $S_0$ and $S_1$ cannot be both codimension-0 subspaces of $v_i$ parameter space, and the codimension-1 gap closing at $\Gamma$ can only be $Z_2$ or non-$Z_2$ but not both.
Since we already show that the $Z_2$ transition at $\Gamma$ can be codimension-1, the codimension-1 gap closing at $\Gamma$ should always change the $Z_2$ index.



We next study the non-zero PET components, starting from the $v_1=v_2=0$ case.
If we further set $\xi_3=\xi_4=\xi_5=\xi_6=0$, the electron-strain coupling $h_{p3,1}$ also has the effective inversion $\widetilde{P}$, leading to the vanishing PET.
It means that $\xi_1$ and $\xi_2$ cannot contribute to the PET for $v_1=v_2=0$.
Indeed, the direct derivation gives the PET jump
\eqa{
\label{eq:PET_p3_Gamma}
&\Delta \gamma_{xxx}=-\frac{e}{\pi}\frac{\sum_{b=3}^6 v_b\xi_b }{\sum_{b=3}^6 v_b^2}\\
&\Delta \gamma_{yyy}=\frac{e}{\pi}\frac{v_4\xi_3-v_3\xi_4+v_6\xi_5-v_5\xi_6}{\sum_{b=3}^6 v_b^2}\ .
}
For non-zero $v_1$ and $v_2$, the PET components can be calculated numerically for $v_3=v_4=v_5=v_6=1 \text{eV}\AA$ and $\xi_1=\xi_2=\xi_3=2\xi_4=\xi_5=2\xi_6=1 \text{eV}$, showing the jump across the TQPT in \figref{fig:SM_Gamma}(a).

\subsubsection{Scenario (ii): $C_3\in\mathcal{G}_0$ and $\mathcal{T}\notin\mathcal{G}_0$}

\label{sec:p3_ii}

The gap closing momenta in scenario (ii) are $K$ and $K'$ in \figGC(g) in the main text.
Since these two momenta are related by $\mathcal{T}$, we only need to derive the effective model at one momentum, say $K$, and the other one can be obtained using $\mathcal{T}$.
At $K$, the $C_3$ symmetry has three possible eigenvalues $-1,\ee^{\pm \ii \pi/3}$ due to $C_3^3=-1$.
If the gap closing is between two states with the same $C_3$ eigenvalues, it cannot be TQPT since the fixed gap closing momentum leads to 3 FTPs for the gap closing. (See \appref{app:p3}.)

There are three cases for two states with different $C_3$ eigenvalues: $(\ee^{-\ii \pi/3},\ee^{\ii \pi/3})$, $(\ee^{\ii \pi/3},-1)$, and $(-1,\ee^{-\ii \pi/3})$.
The effective models in the three cases are equivalent since the representations of $C_3$ in these cases can be related to each other by multiplying a phase factor $\ee^{\pm \ii 2\pi/3}$.
Therefore, we focus on the first case, of which the effective model at $K$ (after an appropriate unitary transformation) is given by $h_+$ in \eqnref{eq:h_+-_p1} with
\eqa{
&v_x=v_y\equiv v,\ v_0=0,\ \xi_{a'',xy}=\xi_{a'',yx}=0, \\
&\xi_{a'',xx}=\xi_{a'',yy},\ \xi_{x,xx}=-\xi_{x,yy}=-\xi_{y,xy}=-\xi_{y,yx},\\
&\xi_{y,xx}=-\xi_{y,yy}=\xi_{x,xy}=\xi_{x,yx}\ ,
}
where  $a''=0 \text{ or } z$.
Similarly, by choosing the appropriate bases at $K'$ such that $\mathcal{T}\dot{=}i\sigma_y\mathcal{K}$, the effective model at $K'$ is given $h_-$ in \eqnref{eq:h_+-_p1} with the parameter relation listed above.
As a result, the gap closing between states with different $C_3$ eigenvalues needs 1 FTP and changes the $Z_2$ index since half 1BZ contains one Dirac cone ($K$ or $K'$).
Based on \eqnref{eq:PET_jump_p1}, the jump of independent PET components across this TQPT (varying $m$ from $0^-$ to $0^+$) has the non-zero form
\eqa{
\label{eq:PET_jump_p3_ii}
&\Delta\gamma_{xxx}=-e\frac{\Delta N_+\xi_{y,xx}}{\pi v}\ ,\ \Delta\gamma_{yyy}=e\frac{\Delta N_+\xi_{x,yy}}{\pi v} \ ,
}
where $\Delta N_+=-\sgn{v^2}=-1$.

\subsubsection{Scenario (iii): trivial $\mathcal{G}_0$}

\label{sec:p3_iii}

In scenario (iii), there are six gap closing momenta, labeled as $\pm \bsl{k}_0$, $\pm C_3\bsl{k}_0$ and $\pm C_3^2\bsl{k}_0$, as shown by red crosses in \figGC(h) in the main text.
The effective Hamiltonian at $\pm \bsl{k}_0$ are exactly the same as \eqnref{eq:h_+-_p1} since the two momenta are related by $\mathcal{T}$ and no more symmetries are involved.
Therefore, the gap closing scenario needs 1 FTP, and the contribution to the PET jump from the gap closing at $\pm\bsl{k}_0$ is the same as \eqnref{eq:PET_jump_p1}, noted as $\Delta \gamma^{(0)}_{ijk}$.
The effective models at $\pm C_3\bsl{k}_0$ and $\pm C_3^2\bsl{k}_0$ can be obtained from those at $\pm \bsl{k}_0$ by $C_3$ and $C_3^2$ operations, respectively, whose electronic parts are also in the Dirac Hamitlonian form.
The contracted half 1BZ then contains three Dirac cones at the gap closing and its CN must change by an odd number, indicating the change of $Z_2$ index.
Furthermore, the contributions to the jump of PET components from the gap closing at $\pm C_3 \bsl{k}_0$ and $\pm C_3^2 \bsl{k}_0$ are $\Delta \gamma^{(1)}_{ijk}=(C_3)_{ii'}(C_3)_{jj'}(C_3)_{kk'}\Delta \gamma^{(0)}_{i'j'k'}$ and $\Delta \gamma^{(2)}_{ijk}=(C_3^2)_{ii'}(C_3^2)_{jj'}(C_3^2)_{kk'}\Delta \gamma^{(0)}_{i'j'k'}$, respectively, owing to the symmetry.
As a result, the jump of independent PET components is given by $\Delta\gamma_{ijk}=\Delta\gamma^{(0)}_{ijk}+\Delta\gamma^{(1)}_{ijk}+\Delta\gamma^{(2)}_{ijk}$, which has the nonzero form
\eqa{
\label{eq:PET_jump_P3_iii}
\Delta\gamma_{xxx}=&-e\frac{3 \Delta N_+}{4 \pi }\frac{2 v_y \xi_{x,xy}-2 v_0 \xi_{y,xy}+v_x(\xi_{y,xx}-\xi_{y,yy}) }{v_x v_y}\\
\Delta\gamma_{yyy}=&e\frac{3 \Delta N_+}{4 \pi}\left[\frac{ v_y (-\xi_{x,xx}+\xi_{x,yy})+2 v_x \xi_{y,xy}}{v_x v_y}\right. \\
&\left. +\frac{v_0(\xi_{y,xx}-\xi_{y,yy})}{v_x v_y}\right]\ .
}

\subsection{PG $p31m$ and PG $p3m1$}

\label{sec:p3m}

Both PGs $p31m$ and $p3m1$ are generated by the lattice translation, the three-fold rotation $C_3$, and a mirror symmetry which we choose to be $m_x$ without loss of generality.
The difference between the two PGs lies on the direction of the mirror line relative to the primitive lattice vector: the mirror line is parallel or perpendicular to one primitive lattice vector for $p31m$ or $p3m1$, respectively.
$C_3$ and $m_x$ span the point group $C_{3v}$, which leads to
\eqa{
\label{eq:PET_sym_p3m}
& \gamma_{xxx}=\gamma_{xyy}=\gamma_{yxy}=\gamma_{yyx}=0\\
& \gamma_{xyx}=\gamma_{xxy}=\gamma_{yxx}=-\gamma_{yyy}
}
for the PET, and thus $\gamma_{yyy}$ serves as the only independent symmetry-allowed PET component.
We classify the gap closing scenarios into 4 types according to $\mathcal{G}_0$: (i) $\mathcal{G}_0$ contains $\mathcal{T}$, (ii) $\mathcal{G}_0$ contains at least one of the three mirror symmetry operations in $C_{3v}$ (again labeled as $\mathcal{U}=m_x$, $C_3 m_x$, or $C_3^2 m_x$) but no $\mathcal{T}$, (iii) $\mathcal{G}_0$ contains the $\mathcal{U}\mathcal{T}$ but no $\mathcal{T}$, and (iv) $\mathcal{G}_0$ is trivial.
As summarized in \tabmainresults in the main text, all gap closing cases between gapped states with 1 FTP change either $Z_2$ index or the VCN, and lead to the jump of symmetry-allowed PET components.

\subsubsection{Scenario (i): TRIM}

\label{sec:p3m_i}

Similar as \secref{sec:p3_i} for PG $p3$, there are four inequivalent TRIM: the $\Gamma$ point and three $M$ points.
Although $\mathcal{G}_0$ at the $M$ point now contains $\mathcal{U}$, the gap closing still requires 3 FTPs same as the corresponding case in \secref{sec:pm_i}, which cannot be a TQPT.

When the gap closes at $\Gamma$ point (\figGC(i-j) in the main text), the generators of $\mathcal{G}_0$ besides the lattice translation are $C_3$, $m_x$ and $\mathcal{T}$, and there are still two types of Kramers pairs characterized by the $C_3$ eigenvalues as those in \secref{sec:p3_i}.
Owing to the extra mirror symmetry here, the number of FTPs for the gap closing between the same type of Kramers pairs becomes $2$ for $(\ee^{-\ii \pi/3},\ee^{\ii \pi/3})$ type and $3$ for $(-1,-1)$ type as discussed in \appref{app:p3m}.
Therefore, we still only need to consider the gap closing between different types of Kramers pairs.
For the convenience of the later material discussion, we choose the bases as $(\ee^{- \ii \pi/3},-1,\ee^{ \ii \pi/3},-1)$.
One can always choose the TR symmetry and mirror symmetry to be represented as $\mathcal{T}\dot{=}-\ii \sigma_y\tau_0\mathcal{K}$ and $m_x\dot{=}-\ii \sigma_x\tau_0$.
In this case, the effective Hamiltonian can be derived by imposing the $m_x$ on \eqnref{eq:hp3_i}, leading to \eq{
\label{eq:p3m_Gamma_cons}
v_1=v_4=v_5=\xi_3=\xi_6=0\ .
}
The form of the Hamiltonian then reads
\eqa{
\label{eq:h_p3m_i}
&h_{p3m}(\bsl{k},u_{ij})=E_0+\xi_{1} u^2 \\
&+\left(
\begin{array}{cccc}
 m & v_3 k_+  & \ii v_{2}k_-  & -\ii v_6 k_+ \\
 v_3 k_-  & -m & -\ii v_6 k_+ & 0 \\
 -\ii v_{2} k_+ & \ii v_6 k_-  & m & -v_3 k_- \\
 \ii v_6 k_-  & 0 & -v_3 k_+ & -m \\
\end{array}
\right)\\
&+
\left(
\begin{array}{cccc}
\xi_2 u^2  & \xi_4 u_- & 0  & \ii \xi_5 u_- \\
 \xi_4 u_+ & -\xi_2 u^2 & -\ii \xi_5 u_- & 0 \\
 0 & \ii \xi_5 u_+ &\xi_2 u^2 & \xi_4 u_+ \\
  -\ii \xi_5 u_+ & 0 & \xi_4 u_- & -\xi_2 u^2 \\
\end{array}
\right)\ ,
}
where $u^2=u_{xx}+u_{yy}$, $u_\pm=u_{xx}-u_{yy}\pm\ii (u_{xy}+u_{yx})$.

The above Hamiltonian shows that the gap closing at $\Gamma$ needs only 1 FTP, which is $m$.
As discussed in \appref{app:p3m}, this gap closing cannot drive a gapped phase into a mirror-protected gapless phase, and therefore can separate two gapped states.
Similar to the discussion in \secref{sec:p3_i}, the gap closing changes the $Z_2$ index when tuning $m$ from $0^-$ to $0^+$, indicating a TQPT.
When $v_2=0$, an analytical expression for the jump of independent PET component can be obtained from \eqnref{eq:PET_p3_Gamma} and \eqnref{eq:p3m_Gamma_cons}, which reads
\eq{
\label{eq:PET_jump_p3m_i}
\Delta \gamma_{yyy}=\frac{e}{\pi}\frac{-v_3\xi_4+v_6\xi_5}{v_3^2+v_6^2}\ .
}
With parameter values $v_3=v_6=1 \text{eV}\AA$ and $\xi_1=\xi_2=2\xi_4=\xi_5=1 \text{eV}$, the numerical results (\figref{fig:SM_Gamma}(b)) for non-zero $v_2$ still show a PET jump across TQPT.

\subsubsection{Scenario (ii): $\mathcal{U}\in \mathcal{G}_0$ and $\mathcal{T}\notin\mathcal{G}_0$}

\label{sec:p3m_ii}

Scenario (ii) can be further divided into two classes depending on whether $\mathcal{G}_0$ contains $C_3$.
When $\mathcal{G}_0$ does not contain $C_3$, the gap closing either requries more than 1 FTP or drives the system into a mirror-protected gapless phase with 0 codimension, similar to \secref{sec:pm_ii}.

Only when the gap closes at $K,K'$ for $p31m$, $\mathcal{G}_0$ contains $C_3$.
In this case, $\mathcal{G}_0$ contains the group $C_{3v}$, which has one 2D irreducible representation (IR) and two different 1D IRs when acting on the states.
The gap closing between the states furnishing the same IR requires 3 FTPs, similar to the case for two states with the same $C_3$ eigenvalue in \secref{sec:p3_ii}.
If the gap closes between the doubly degenerate states furnishing the 2D IR and a state furnishing a 1D IR, the system with a fixed carrier density cannot be insulating on both sides of the gap closing because the number of occupied bands is changed.
If the gap closes between two states that furnish different 1D IRs, the mirror-protected gapless phase must exist on one side of the gap closing as the two states must have opposite mirror eigenvalues.
Therefore, there is no direct TQPT between the insulating phases in scenario (ii).

\subsubsection{Scenario (iii): $\mathcal{UT}\in \mathcal{G}_0$ and $\mathcal{T}\notin\mathcal{G}_0$}

\label{sec:p3m_iii}

In scenario (iii), the gap closing cases are again divided into two different classes depending on whether $\mathcal{G}_0$ has $C_3$.
We first discuss the class without $C_3$, which happens for the gap closing at $\mathcal{UT}$ invariant momenta except $K,K'$ for $p3m1$.
As shown in \figGC(k-m) in the main text, the total number of inequivalent gap closing momenta is six, including $\pm \bsl{k}_0$, $\pm C_3 \bsl{k}_0$, and $\pm C_3^2\bsl{k}_0$.
Without loss of generality, we choose $\bsl{k}_0$ such that $-m_x\bsl{k}_0$ is equivalent to $\bsl{k}_0$.
Then, the effective models at $\pm \bsl{k}_0$ are the same as the corresponding models in \secref{sec:pm_iii}, \ie \eqnref{eq:h_+-_p1} with the parameter relation \eqnref{eq:para_pm_iii}, indicating 1 FTP for the gap closing.
Since the effective models at $\pm C_3\bsl{k}_0$ and $\pm C_3^2\bsl{k}_0$ are related to those at $\pm \bsl{k}_0$ by $C_3$ and $C_3^2$ operations similar to \secref{sec:p3_iii}, the jump of PET components can be derived by substituting \eqnref{eq:para_pm_iii} into \eqnref{eq:PET_jump_P3_iii}, resulting in
\eq{
\Delta\gamma_{yyy}=e\frac{3 \Delta N_+}{4 \pi}\frac{ v_y (-\xi_{x,xx}+\xi_{x,yy})+2 v_x \xi_{y,xy}}{v_x v_y}\ .
}
Moreover, since three Dirac cones exist in half 1BZ when the gap closes, the $Z_2$ index changes at the gap closing, making it a TQPT.

The class that $\mathcal{G}_0$ includes $C_3$ can only happen when the gap closes at $K$ and $K'$ for PG $p3m1$, as shown in \figGC(m) in the main text.
We can choose $C_3$ and $m_x\mathcal{T}$ as the generators of $\mathcal{G}_0$ besides the lattice translation.
Similar to \secref{sec:p3_ii}, we first study $K$ and derive the model at $K'$ by choosing the right bases such that $\mathcal{T}\dot{=}\ii \sigma_y\mathcal{K}$.
The states at $K$ can be labeled by $C_3$ eigenvalues, $-1$ and $\ee^{\pm\ii \pi/3}$ given by $C_3^{3}=-1$.
Since $(m_x \mathcal{T})^2=1$ and $C_3 m_x\mathcal{T}=m_x\mathcal{T} C_3^{-1}$, the gap closing typically happens between two non-degenerate states, labeled by the $C_3$ eigenvalues as $(\lambda_1,\lambda_2)$, and we can always choose $m_x\mathcal{T}\dot{=}\mathcal{K}$.
The $\lambda_1=\lambda_2$ case cannot correspond to TQPT since 2 FTPs are needed for the gap closing as discussed in \appref{app:p3m}, while the $\lambda_1\neq \lambda_2$ case requires only one FTP for the gap closing similar to \secref{sec:p3_ii}.
Since the matrix representations of $C_3$ and $m_x\mathcal{T}$ are equivalent for the three choices $(\lambda_1,\lambda_2)=(\ee^{-\ii \pi/3},\ee^{\ii \pi/3}),\ (\ee^{\ii \pi/3},-1),$ and $(-1,\ee^{-\ii \pi/3})$, they have the same effective models and we only consider the first choice.
With all the above conventions and simplifications, the effective models at $K$ and $K'$ can be given by those for \secref{sec:p3_ii} with an extra constraint $\xi_{y,yy}=0$ brought by $m_x\mathcal{T}$.
As a result, the $Z_2$ index does change when the gap closes, and the jump of PET components can be derived from \eqnref{eq:PET_jump_p3_ii} with the above extra constraint, which reads
\eq{
\label{eq:PET_jump_p3m1_K}
\Delta\gamma_{yyy}=-e\frac{\Delta N_+ \xi_{x,xx}}{\pi v} \ .
}

\subsubsection{Scenario (iv): trivial $\mathcal{G}_0$}

\label{sec:p3m_iv}

In scenario (iv), the gap closes simultaneously at twelve inequivalent momenta, namely $\pm \bsl{k}_0$, $\pm m_x\bsl{k}_0$, $\pm C_3\bsl{k}_0$, $\pm C_3 m_x\bsl{k}_0$, $\pm C_3^2\bsl{k}_0$ and $\pm C_3^2 m_x\bsl{k}_0$ in \figGC(n-o) in the main text.
The effective model around $\bsl{k}_0$ can be chosen as $h_+$ in \eqnref{eq:h_+-_p1}, and the models around other gap closing momenta can be further obtained by the symmetry.
Although this gap closing scenario only needs 1 FTP, it cannot induce any change of $Z_2$ index since there is an even number (six) of Dirac cones in contracted half 1BZ.
However, the gap closing can change the VCN when the twelve valleys are well defined according to \appref{sec:pm_iv}, \eg $N_{\bsl{k}_0}$ can change by $\pm 1$, and thus is a TQPT in the sense of the locally stable topology.

We split the change of PET components for this scenario into 3 parts: $\gamma^{(0)}$ from $\pm \bsl{k}_0$ and $\pm m_x\bsl{k}_0$, $\gamma^{(1)}$ from $\pm C_3\bsl{k}_0$ and $\pm C_3 m_x\bsl{k}_0$, and $\gamma^{(2)}$ from  $\pm C_3^2\bsl{k}_0$ and $\pm C_3^2 m_x\bsl{k}_0$.
Since the contribution to $\gamma^{(0)}$ contains two Kramers pairs that are related by $m_x$, same as \secref{sec:pm_iv}, $\gamma^{(0)}$ equals to \eqnref{eq:PET_pm_iv}.
$C_3$ symmetry then gives $\Delta \gamma^{(1)}_{ijk}=(C_3)_{ii'}(C_3)_{jj'}(C_3)_{kk'}\Delta \gamma^{(0)}_{i'j'k'}$ and $\Delta \gamma^{(2)}_{ijk}=(C_3^2)_{ii'}(C_3^2)_{jj'}(C_3^2)_{kk'}\Delta \gamma^{(0)}_{i'j'k'}$, similar to \secref{sec:p3_iii}.
As the result, the total change of PET can be obtained from $\Delta\gamma=\Delta\gamma^{(0)}+\Delta\gamma^{(1)}+\Delta\gamma^{(2)}$,
which is propotional to the change of the VCN of the system
\eqa{
\label{eq:PET_jump_P3m_iv}
\Delta\gamma_{yyy}&=e\frac{ \Delta N_{\text{val}}}{8 \pi}\left[\frac{ v_y (-\xi_{x,xx}+\xi_{x,yy})+2 v_x \xi_{y,xy}}{v_x v_y}\right.\\
&\left.+\frac{v_0(\xi_{y,xx}-\xi_{y,yy})}{v_x v_y}\right]
}
with $\Delta N_{\text{val}}=12 \Delta N_+$.

\subsection{10 PGs with 2D Inversion or $C_2$}
\label{sec:C2}

The PET jump cannot exist in 10 PGs that contain $C_2$ or inversion, including $p2$, $p2mm$, $p2mg$, $p2gg$, $c2mm$, $p4$, $p4mm$, $p4gm$, $p6$, and $p6mm$.
This conclusion can be drawn from the symmetry analysis of PET.
Since both $C_2$ and inversion transform $(x,y)$ to $(-x,-y)$, $\gamma_{ijk}=-\gamma_{ijk}$ is required for those 10 PGs, leading to the vanishing PET.
Early study\cite{Ahn2017C2T,Fang2015TCI} also shows that a stable gapless phase can exist in between the QSH insulator and the NI when $C_2$ exists.
In this gapless regime, 2D gapless Dirac fermions are locally stable and can only be created or annihilated in pairs.

\section{Numbers of FTPs and Effective Models for the Gap Closing}
The discussion on the gap closing between two non-degenerate states has some overlap with \refcite{Yang20172DGC}.

\subsection{PG $p1$}
\label{app:p1}

This part has been studied in \refcite{Murakami2007QSH}.

\subsubsection{Not TRIM}

When the gap closes at $\bsl{k}_0$ that is not a TRIM, the two-band model near the gap closing to the leading order of $\bsl{q}=\bsl{k}-\bsl{k}_0$ in general takes the form
\eqa{
\label{eq:h_p1_app}
h(\bsl{q})=E_0(\bsl{q})\sigma_0+( q_x \bsl{C}_{x}+q_y \bsl{C}_{y}+\bsl{M})\cdot \boldsymbol{\sigma}\ ,
}
where $\bsl{C}_i=(C_{ix}, C_{iy}, C_{iz})$, $\bsl{M}=(M_x,M_y,M_z)$, $\boldsymbol{\sigma}=(\sigma_x,\sigma_y,\sigma_z)$, and the two bases of the above model account for the doubly degenerate band touching when the gap closes.
\eqnref{eq:h_p1_app} determines the codimension for the gap closing scenario since the gap at $-\bsl{k}_0$ is related to that of \eqnref{eq:h_p1_app} by the TR symmetry.
The gap of \eqnref{eq:h_p1_app} closes if and only if $\bsl{C}_x q_x +\bsl{C}_y q_y +\bsl{M}=0$.
We choose $\bsl{C}_x\times \bsl{C}_y\neq 0$ since it can be satisfied without finely tuning anything (or equivalently in a parameter subspace with 0 codimension).
In this case, the gap closes when $\bsl{M}$ lies in the plane spanned by two vectors $\bsl{C}_x$ and $\bsl{C}_y$.
Therefore, the codimension for the gap closing is 1 since only the angle between the vector $\bsl{M}$ and the $(\bsl{C}_x,\bsl{C}_y)$ plane needs to be tuned.

Next, we derive \eqhpsim of the main text and the electronic part of $h_+$ in \eqnref{eq:h_+-_p1}, while the model at $-\bsl{k}_0$ can be derived by the TR symmetry and thus is not discussed here.
\eqnref{eq:h_p1_app} always allows the $\bsl{q}$-independent $\SU(2)$ transformation, \ie $h(\bsl{q})\rightarrow U^\dagger h(\bsl{q}) U$ with $U\in \SU(2)$.
Such transformation only changes the bases of the Hamiltonian but does not change the direction of the momentum or the coordinate system.
Since $\boldsymbol{\sigma}$ behaves as an $\SO(3)$ vector under $U$, every $\SU(2)$ transformation $U$ of the Hamiltonian is equivalent to an $\SO(3)$ transformation $R$ of the vectors $\bsl{C}_i$ and $\bsl{M}$, \ie $\bsl{C}_i\rightarrow R\bsl{C}_i$ and $\bsl{M}\rightarrow R\bsl{M}$.
Thus, by choosing appropriate $U$ matrix, we can first rotate $\bsl{C}_x$ to the $x$ direction and then $\bsl{C}_y$ to the $xy$ plane, resulting in $R\bsl{C}_x=v_x \bsl{e}_x$, $R\bsl{C}_y=v_0 \bsl{e}_x+v_y\bsl{e}_y$ and $R\bsl{M}=m_1\bsl{e}_x+m_2\bsl{e}_y+m\bsl{e}_z$.
As a result, \eqnref{eq:h_p1_app} is transformed to
\eqa{
h(\bsl{q})=&E_0(\bsl{q})\sigma_0+(v_x q_x+v_0 q_y+m_1)\sigma_x+(v_y q_y+m_2)\sigma_y\\
& + m \sigma_z\ .
}
Here $\bsl{C}_x\times\bsl{C}_y\neq 0$ gives non-zero $v_x$ and $v_y$.
With a shift of $\bsl{k}_{0}$ by $(m_1/v_x- v_0 m_2/(v_x v_y), m_2/v_y)$, the model is further simplified to the electronic part of $h_+$ in \eqnref{eq:h_+-_p1}.
Finally, we define the $q_x+v_0 q_y$ and $q_y$ to be $q_1$ and $q_2$, respectively, to get \eqhpsim, which represents the most generic form of the Hamiltonian.

\subsubsection{TRIM}
In this part, we count the number of FTPs for the gap closing at the TRIM.
Owing to the Kramers' degeneracy, every band at the TRIM is doubly degenerate, and we use the name "Kramers pair" to label the two states related by the TR symmetry.
We consider the gap closing between two Kramers pairs $\ket{1,\pm}$ and $\ket{2,\pm}$, where $\TR \ket{i,+}=-\ket{i,-}$ can always be chosen by the unitary transformation.
As a result, the mass term for the effective model at the TRIM reads
\eq{
\label{eq:TRIM_p1}
\mat{
m_1 & 0 & \Delta_0+\ii \Delta_3 & \ii \Delta_1+\Delta_2 \\
0 & m_1 & \ii \Delta_1-\Delta_2  & \Delta_0-\ii \Delta_3\\
\Delta_0-\ii \Delta_3 & -\ii \Delta_1-\Delta_2 & m_2 & 0\\
-\ii \Delta_1+\Delta_2 & \Delta_0+\ii \Delta_3 & 0 & m_2 \\
}
}
where the bases are $(\ket{1,+},\ket{1,-},\ket{2,+},\ket{2,-})$ and all parameters are real.
Since the momentum is fixed at TRIM ($-\bsl{k}=\bsl{k}+\bsl{G}$ with $\bsl{G}$ a reciprocal lattice vector), none of the terms in the above equation can be canceled by shifting the momentum.
Therefore, 5 FTPs are needed for the gap closing.

\subsection{$p1m1$, $c1m1$, and $p1g1$}
\label{app:pm}
\subsubsection{Scenario (i): TRIM}

If $\mathcal{G}_0$ does not contain $\mathcal{U}$, which can occur on the edge of 1BZ for $c1m1$, the situation is the same as the TRIM in \appref{app:p1}, which requires 5 FTPs.
When $\mathcal{G}_0$ contains $\mathcal{U}$, we should discuss the $\mathcal{U}=m_x$ case ($p1m1$ and $c1m1$) and the $\mathcal{U}=g_x$ case ($p1g1$), separately.

For $p1m1$ and $c1m1$, since $m_x^2=-1$, two states of one Kramers pair have opposite mirror eigenvalues $\pm \ii$, labeled by $m_x\ket{i,\pm}=\pm \ii \ket {i,\pm}$.
On the bases $(\ket{1,+},\ket{1,-},\ket{2,+},\ket{2,-})$, the effective model around the gap closing between two Kramers pairs can be given by \eqnref{eq:TRIM_p1} with $\Delta_1=\Delta_2=0$, since the bases with different mirror eigenvalues cannot be coupled by the mass terms.
As a result, 3 FTPs are needed for such gap closing scenario.

For $p1g1$, $g_x^2=-1$ at $\Gamma$ and $X$ and the number of FTPs for the gap closing is thus the same as the above case, which is 3.
At $Y$ and $M$, $g_x^2=1$ and two states of one Kramers pair have the same $g_x$ eigenvalue, $1$ or $-1$.
In this case, the gap closing between two Kramers pairs with the same $g_x$ eigenvalue needs 5 FTPs, which is the same as the TRIM scenario in \appref{app:p1}.
On the other hand, between two Kramers pairs with opposite $g_x$ eigenvalues, only 1 FTP needs to be tuned to close the gap at $Y$ or $M$, since the off-diagonal terms ($\Delta_{0,1,2,3}$) in \eqnref{eq:TRIM_p1} are all forbidden.

\subsubsection{Scenario (ii): $\mathcal{U}\in \mathcal{G}_0$ but $\TR\notin \mathcal{G}_0$}

In scenario (ii), there are two gap closing momenta $\pm\bsl{k}_0$ that are related by the TR symmetry.
Therefore, we only need to consider one of them, say $\bsl{k}_0$, to derive the number of FPTs for the gap closing.
At $\bsl{k}_0$, the states can be labeled by the eigenvalues of $\mathcal{U}$.
If the gap closing between two states with the same $\mathcal{U}$ eigenvalues, the effective model can be described by \eqnref{eq:h_p1_app} with $|\bsl{C}_x|=0$. 
The gap closes if and only if $\bsl{C}_y q_y+\bsl{M}=0$, realizable by making two vectors $\bsl{M}$ and $\bsl{C}_y$ parallel.
Such realization needs 2 FTPs, \eg the two components of the projection of $\bsl{M}$ on the plane perpendicular to $\bsl{C}_y$.

When the gap closes between two states with different $\mathcal{U}$ eigenvalues, the effective model along the $\mathcal{U}$-invariant line ($q_x=0$) reads
$
h({\bf q})=E_0(q_y)+(m_0+C q_y+B_0 q_y^2)\sigma_z\ ,
$
which, by shifting the $k_{0,y}$, can be simplified to $h({\bf q})=E_0(q_y)+(m+B q_y^2) \sigma_z$.
The gap for this Hamiltonian keeps closing when $m B\leq 0$, indicating a stable gapless phase protected by $\mathcal{U}$ with 0 codimension.

\subsubsection{Scenario (iii): $\mathcal{U}\TR\in \mathcal{G}_0$ but $\TR\notin \mathcal{G}_0$}

In this scenario, we here only consider the $(\mathcal{U}\TR)^2=-1$ case, where each band at the gap closing momentum is doubly degenerate.
We can define the $\mathcal{U}\TR$ pair as the two degenerate states related by $\mathcal{U}\TR$, in analog to the Kramers pair defined in \appref{app:p1}.
Similar as \eqnref{eq:TRIM_p1}, there are 5 mass terms for the gap closing between two $\mathcal{U}\TR$ pairs.
However, the case here is different from the TRIM scenario in \appref{app:p1}, since $q_x$ does not change under $\mathcal{U}\TR$ and thus the corresponding terms have the same form as the mass terms in \eqnref{eq:TRIM_p1}.
One of the five mass terms can then be canceled by shifting $k_{0,x}$, resulting in 4 FTPs for the gap closing.

\subsection{$p_3$}
\label{app:p3}

\subsubsection{Scenario (i):TRIM}

We first discuss the gap closing at $\Gamma$ between two Kramers pairs of the same type.
If the bases have $C_3$ eigenvalues $(\ee^{-\ii\pi/3},\ee^{\ii \pi/3},\ee^{-\ii\pi/3},\ee^{\ii \pi/3})$, the mass term of the effective model is given by \eqnref{eq:TRIM_p1} with $\Delta_1=\Delta_2=0$ since the bases with different $C_3$ eigenvalues cannot be coupled, resulting in 3 FTPs for the gap closing.
If the bases have $C_3$ eigenvalues $(-1,-1,-1,-1)$, the effective model equals to \eqnref{eq:TRIM_p1} that has 5 FTPs for the gap closing.

Now we discuss the construction of the effective model for the bases $(\ee^{- \ii \pi/3},\ee^{ \ii \pi/3},-1,-1)$.
The form of the effective model, \eqnref{eq:hp3_i}, is given by the tensor product of the bases in the same IR listed in \tabref{tab:P3_Sym}.
Note that the matrix representation and the bases for the $E$ IR are not Hermitian.
It means given two copies of $(E,+)$ or $(E,-)$ IR, say $(\tau_+(\sigma_x-\ii\sigma_y),\tau_+(\sigma_x+\ii\sigma_y))$ and $(k_x-\ii k_y,k_x+\ii k_y)$ furnishing $(E,-)$ IR, the coefficients used for the tensor product can be complex, \eg $c[\tau_+(\sigma_x-\ii\sigma_y)][k_x-\ii k_y]^*+c^*[\tau_+(\sigma_x+\ii\sigma_y)][k_x+\ii k_y]^*$ with complex $c$.

\begin{table*}[t]
    \centering
    \begin{tabular}{cc}
    \hline
    IR & Expressions\\
    \hline
    \hline
         $A_1,+$ &  $\tau_+\sigma_0$, $\tau_-\sigma_0$, $u_{11}+u_{22}$  \\
         $A_1,-$ &  $\tau_+\sigma_z$, $\tau_-\sigma_x$, $\tau_-\sigma_y$, $\tau_-\sigma_z$  \\
         $E,+$ & $(\tau_y\sigma_z-\ii \tau_x\sigma_0,\tau_y\sigma_z+\ii \tau_x\sigma_0)$, $(\tau_y\sigma_x+\ii \tau_y\sigma_y,\tau_y\sigma_x-\ii \tau_y\sigma_y)$, $(-u_{xx}+u_{yy}-\ii (u_{xy}+u_{yx}),-u_{xx}+u_{yy}+\ii (u_{xy}+u_{yx}))$\\
         $E,-$ & $(\tau_+(\sigma_x-\ii\sigma_y),\tau_+(\sigma_x+\ii\sigma_y))$, $(\tau_x\sigma_z+\ii\tau_y\sigma_0,\tau_x\sigma_z-\ii\tau_y\sigma_0)$, $(\tau_x\sigma_x+\ii\tau_x\sigma_y,\tau_x\sigma_x-\ii\tau_x\sigma_y)$, $(k_x-\ii k_y,k_x+\ii k_y)$\\
         \hline
    \end{tabular}
    \caption{The irreducible representations (IRs) of $C_{3}$ and TR symmetries.
    In $A_1$ IR, the $C_3$ eigenvalue of the bases is $1$ and $\pm$ are parity under TR.
    ``$E,\pm$" label two 2D IRs, where the two components have the $C_3$ eigenvalues $(\ee^{\ii 2\pi/3},\ee^{-\ii 2\pi/3})$ and transform as $\pm \sigma_x\cc$ under the TR symmetry.
    $\tau_\pm=(\tau_0\pm \tau_z)/2$.
    \label{tab:P3_Sym}
    }
\end{table*}

\subsubsection{Scenario (ii): $C_3\in\mathcal{G}_0$ and $\mathcal{T}\notin\mathcal{G}_0$}

Here we consider the gap closing between two states with the same $C_3$ eigenvalues at $K$ or $K'$.
In general, the mass terms at one gap closing momentum are $m_x \sigma_x+m_y \sigma_y+m_z \sigma_z$.
Since the gap closing momentum is fixed, none of the three mass terms can be canceled by shifting the momentum, and hence there are 3 FTPs for the gap closing.

\subsection{$p31m$ and $p3m1$}
\label{app:p3m}

\subsubsection{Scenario (i): TRIM}

When the two Kramers pairs carry $C_3$ eigenvalues as $(\ee^{-\ii\pi/3},\ee^{\ii \pi/3},\ee^{-\ii\pi/3},\ee^{\ii \pi/3})$, the effective model equals to \eqnref{eq:TRIM_p1} with $\Delta_1=\Delta_2=0$ before considering $m_x$, similar to the correspond case in \appref{app:p3}.
As $m_x\dot{=}-\ii \sigma_x$ for each Kramers pair, the $\Delta_3$ is also forbidden, resulting in 2 FTPs for the gap closing.
On the other hand, if $C_3$ eigenvalues are all $-1$, the effective model equals to \eqnref{eq:TRIM_p1} before considering $m_x$, similar to the correspond case in \appref{app:p3}, and including $m_x$ makes $\Delta_2=\Delta_3=0$, leading to 3 FTPs for the gap closing.

The construction of the effective model for the bases $(\ee^{- \ii \pi/3},-1,\ee^{ \ii \pi/3},-1)$ is the same as that for the (111) HgTe/CdTe quantum well, which is discussed in \appref{app:HgTe}.
Next we show that the gap closing at $\Gamma$ in this case cannot drive a gapped system to the mirror protected gapless phase.
Since the three mirror lines are related by the $C_3$ symmetry, we only need to consider one of them, say $k_x=0$ that is invariant under $m_x$.
The eigenvalues along this line read
\eq{
E_{\alpha\beta}=E_0+\alpha \frac{k_y v_2}{2}+\beta \sqrt{(m+\alpha \frac{v_2 k_y}{2})^2+k_y^2 (v_3^2+v_6^2)}
}
with $\alpha,\beta$ take $\pm$.
$E_{\pm\beta}$ bands cross at $\Gamma$ and belong to the same set of connected bands.
The mirror eigenvalue of the $E_{\alpha\beta}$ band is $-\alpha\ii$, and then the mirror protected gapless phase happens when $E_{++}$ crosses with $E_{--}$ or $E_{+-}$ crosses with $E_{-+}$.
Both band crossings require the same condition
\eq{
|v_2 k_y|=\sum_\alpha\sqrt{(m+ \alpha\frac{v_2 k_y}{2})^2+k_y^2 (v_3^2+v_6^2)},
}
 since they are related by the TR symmetry.
However, the above equation has no solution when $m\neq 0$ and $v_3^2+v_6^2\neq 0$.
It can be seen from the inequality $\sqrt{(a+b)^2+c^2}+\sqrt{(a-b)^2+c^2}> 2 |b|$, which holds unless $c=0$ and $|a|\leq |b|$.
Therefore, without finely tuning more parameters to realize $v_3^2+v_6^2=0$, a gapped system remains when the sign of $m$ flips.

\subsubsection{Scenario (iii): $\mathcal{UT}\in \mathcal{G}_0$ and $\mathcal{T}\notin\mathcal{G}_0$}
Here we discuss the case when the gap closes at $K$ and $K'$ for PG $p3m1$ and between two states with the same $C_3$ eigenvalues.
Before considering $m_x\TR$, the mass terms at $K$ are $m_x\sigma_x+m_y\sigma_y+m_z\sigma_z$ since $C_3$ does not provide any constraints and the fixed gap closing momentum cannot be shifted to cancel any of them.
Since $m_x\TR$ can be chosen as $\sigma_0\cc$, $m_y$ is forbideen and the remaining two mass terms serve as the 2 FTPs for the gap closing.

\section{VCN in Tight-Binding Model}

\begin{figure*}[t]
\includegraphics[width=0.8\columnwidth]{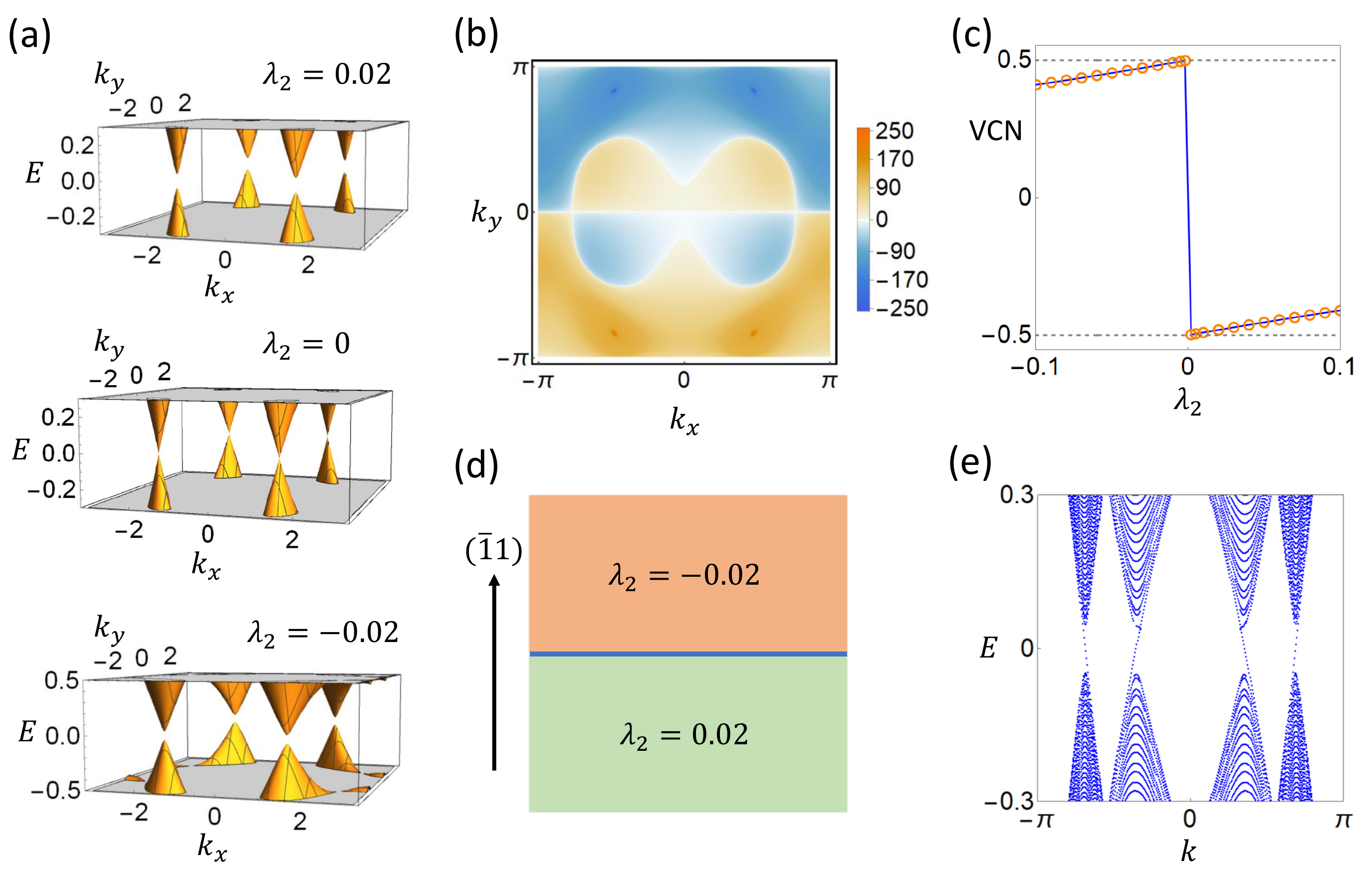}
\caption{
(a) shows the energy dispersion of the TB model \eqnref{eq:h_TB_p1m1} for $\lambda_2=0.02,0,-0.02$. 
The gap is zero for $\lambda_2=0$.
(b) shows the distribution of the Berry curvature for $\lambda_2=0.02$.
The peaks indicate the locations of valleys.
(c) shows the integration of Berry curvature divided by $2\pi$ over the $k_{x,y}>0$ quarter of 1BZ.
The orange circles are the data points, based on which the blue line is plotted.
(d) is a domain wall structure  along $(\bar{1}1)$ direction.
The lower and upper parts are given by the TB models for $\lambda_2=-0.02$ and $\lambda_2=0.02$, respectively.
(e) plots the energy dispersion of domain-wall modes (modes with considerable probability near the interface in (d)).
Here we choose the number of unit cells along $(\bar{1}1)$ to be 70 for each part of the domain wall. 
}
\label{fig:p1m1_TB}
\end{figure*}

In this section, we discuss the quantization and physical meaning of the VCN change in a tight-binding (TB) model with $p1m1$.
We consider a square lattice and each unit cell only contains one atom.
Without loss of generality, we set the lattice constant to 1, and choose the mirror symmetry as $m_y$.
On each atom, we include a spinful $s$ and a spinful $p_y$ orbitals, meaning that the bases can be labeled as $\ket{\bsl{R},\alpha,s}$ with $\bsl{R}$ the lattice vector, $\alpha=s,p_y$ for orbital, and $s=\uparrow,\downarrow$ for spin.
The bases with specific Bloch momentum can be obtained by the following Fourier transformation
\eq{
\ket{\bsl{k},\alpha,s}=\frac{1}{\sqrt{N}}\sum_{\bsl{k}}\ee^{\ii \bsl{k}\cdot\bsl{R}}\ket{\bsl{R},\alpha,s}\ .
}
Then, the representations of the symmetries read
\eqa{
&m_y\ket{\bsl{k},\alpha,s}=\ket{m_y\bsl{k},\alpha',s'}(\tau_z)_{\alpha'\alpha}(-\ii \sigma_y)_{s's}\\
&\TR\ket{\bsl{k},\alpha,s}=\ket{-\bsl{k},\alpha,s'}(\ii \sigma_y)_{s's}\ ,
}
where $\tau$ and $\sigma$ are Pauli matrices for orbital and spin.

With on-site terms and nearest-neighbor hopping terms, we choose the following symmetry-allowed expression for the Hamiltonian
\eq{
\label{eq:h_TB_p1m1}
h(\bsl{k})=d_1 \tau_z\sigma_0+d_2 \tau_y\sigma_z+d_3\tau_x\sigma_z+d_4\tau_y\sigma_0\ ,
}
where 
\eqa{
& d_1=m+2 t_1 \cos(k_x)+2 t_2 \cos(k_y)\\
& d_2=\lambda_1\\
& d_3=2\lambda_2\sin(k_x)\\
& d_4=2 t_3 \sin(k_y)\ .
}
The eigenvalues of $h(\bsl{k})$ read $\pm\sqrt{d_1^2+d_3^2+(d_2\pm d_4)^2}$.
For concreteness, we choose $t_1=t_2=t_3=\lambda_1=1/2$ and $m=4/5$ and assume the model is half-filled (two occupied bands).
In this case, the gap closes only when we tune $\lambda_2$ to zero, as shown in \figref{fig:p1m1_TB}(a), and the gap closing points sit at $\bsl{k}=(\pm \arccos(\frac{-8+5\sqrt{3}}{10}),\pm 5\pi/6)\approx (\pm 1.50, \pm 2.62)$, belonging to the VCN scenario (iv) for $p1m1$.
When the gap is small but nonzero, the positions of valleys can be determined numerically by locating the peaks of Berry curvature (\figref{fig:p1m1_TB}(b)), which are close to the gap closing points.

Finally, based on the TB model \eqnref{eq:h_TB_p1m1}, we discuss the quantization and physical consequence of the VCN change across the $\lambda_2=0$ gap closing.
Without loss of generality, we take the valley in the $k_{x,y}>0$ quarter of 1BZ as an example to discuss the quantization.
The VCN of this valley can be calculated by integrating the Berry curvature over the $k_{x,y}>0$ quarter of 1BZ.
As shown in \figref{fig:p1m1_TB}(c), although VCN is not quantized on any side of the gap closing, the change of VCN across the gap closing is an integer, consistent with the effective-model analysis in the main text.
According to \refcite{Martin2010VCN}, one physical consequence of the quantized VCN change is the gapless domain-wall mode in a domain wall structure that consists of the two different gapped phases separated by the gap closing, like \figref{fig:p1m1_TB}(d).
As shown in \figref{fig:p1m1_TB}(e), the VCN change for each valley matches the number of gapless domain-wall modes around that valley.

\section{(111) HgTe/CdTe Quantum Well}

\label{app:HgTe}

In this section, we provide more details on the analysis of the HgTe QW.
Before going into details, we first introduce some basic properties of the QW.
Both HgTe and CdTe have the standard zinc-blende structure, similar to most II-VI or III-V compound semiconductors.
The crystallographic space group of both compounds is $F\bar{4}3m$ (space group No.\,216).
In the QW, HgTe serves as a well while Hg$_{1-x}$Cd$_{x}$Te serves as the barrier.
Similar to early experimental and theoretical studies~\cite{Bernevig2006BHZ,Konig2007QSHHgTe,Li2009HgTeEfield,
Rothe2010HgTe,Novik2005HgMnTe}, we use $x=0.7$.

\subsection{$d$-induced PET jump for $\mathcal{E}=0$}

To describe the TPQT, we project the 6-band Kane model onto the bases $(\ket{E_1,+},\ket{H_1,+},\ket{E_1,-},\ket{H_1,-})$ via second order perturbation~\cite{Rothe2010HgTe} and get the following 4-band model
\eqa{
\label{eq:heff_0_HgTe_E0}
&h_{eff}^{(0)}(\mathcal{E}=0)=E_0+ B_0 k_\shpa^2\\
&+
\left(
\begin{array}{cccc}
 B k_{\shpa}^2+m & A_{1} k_+ & 0  & -\ii A_{2} k_+ \\
 A_{1} k_- & -B k_{\shpa}^2-m & -\ii A_{2} k_+ & 0 \\
 0& \ii A_{2} k_- & B k_{\shpa}^2+m & -A_{1} k_- \\
 \ii A_{2} k_- & 0 & -A_{1} k_+ & -B k_{\shpa}^2-m \\
\end{array}
\right)\ ,
}
where the values of the parameters are listed \tabref{tab:para_val_eff_E0}, $k_\shpa^2=k_1^2+k_2^2$, $k_\pm=k_1\pm\ii k_2$, and $k_1$ and $k_2$ are the momenta along $(1,-1,0)$ and $(1,1,-2)$, respectively.
Compared to the celebrated Bernevig-Hughes-Zhang model~\cite{Bernevig2006BHZ}, we have an additional $k$-linear term $A_2$ due to the reduction of the full rotational symmetry to $C_3$ rotation symmetry.
In the \eqnref{eq:heff_0_HgTe_E0}, the TQPT shown in \figHgTe(a) of the main text occurs at $m=0$.
To show the jump of the symmetry-allowed PET components at the gap closing, we need to introduce the electron-strain coupling $h_{eff}^{(1)}$ based on the symmetry:
\eqa{
\label{eq:heff_1_HgTe}
h_{eff}^{(1)}=\xi_{1} u^2+
\left(
\begin{array}{cccc}
\xi_2 u^2  & \xi_3 u_- & 0  & -\ii \xi_4 u_- \\
 \xi_3 u_+ & -\xi_2 u^2 & \ii \xi_4 u_- & 0 \\
 0 & -\ii \xi_4 u_+ &\xi_2 u^2 & \xi_3 u_+ \\
  \ii \xi_4 u_+ & 0 & \xi_3 u_- & -\xi_2 u^2 \\
\end{array}
\right)\ ,
}
where $u^2=u_{11}+u_{22}$ and $u_\pm=u_{11}-u_{22}\pm\ii (u_{12}+u_{21})$.
This electron-strain coupling is in the most general symmetry-allowed form to the leading order of $u_{ij}$, which definitely includes the IB terms, $\xi_3$ and $\xi_4$.
With \eqnref{eq:heff_0_HgTe_E0} and \eqnref{eq:heff_1_HgTe}, the independent PET component $\gamma_{222}$ can be derived analytically as
\eqa{
\label{eq:gamma_222_HgTe}
& \gamma_{222}=\\
&-e\frac{\text{sgn}(m) (A_{1} \xi_{3}+A_{2} \xi_{4})}{2 \pi  \left(A_{1}^2+A_{2}^2\right)}\frac{ \left(A_{1}^2+A_{2}^2-2 \left| B m\right| +2 B m\right)}{ \left(A_{1}^2+A_{2}^2+4 B m\right)}\ ,
}
resulting in the PET jump as
\eq{
\Delta\gamma_{222}=-e\frac{A_{1} \xi_{3}+A_{2} \xi_{4}}{ \pi  \left(A_{1}^2+A_{2}^2\right)}\ .
}
Based on \eqnref{eq:gamma_222_HgTe} and $\xi_{1,2,3,4}=1$eV (comparable to those in \refcite{Rostami2018PiE}), we plot the $\gamma_{222}$ of the function of the width in \figHgTe(b) of the main text, which shows a jump around $d=65\AA$.

\subsection{$\mathcal{E}$-induced PET jump for fixed $d$}

After including the electric field, we project the modified Kane model onto the bases $(\ket{E_1,+},\ket{H_1,+},\ket{E_1,-},\ket{H_1,-})$ via second order perturbation and get the following 4-band model
\eq{
\label{eq:heff_0_HgTe}
h_{eff}^{(0)}=E_0+ B_0 k_\shpa^2+
\left(
\begin{array}{cccc}
 B k_{\shpa}^2+m & A_{1} k_+ +D_{1} k_-^2 & -\ii D_{3}k_-  & -\ii A_{2} k_+-\ii D_{2} k_-^2 \\
 A_{1} k_- +D_{1} k_+^2 & -B k_{\shpa}^2-m & -\ii A_{2} k_+ +\ii D_{2} k_-^2 & 0 \\
 \ii D_{3} k_+ & \ii A_{2} k_- -\ii D_{2} k_+^2 & B k_{\shpa}^2+m & D_{1} k_+^2-A_{1} k_- \\
 \ii A_{2} k_- + \ii D_{2} k_+^2 & 0 & D_{1} k_-^2-A_{1} k_+ & -B k_{\shpa}^2-m \\
\end{array}
\right)\ .
}
Compared with \eqnref{eq:heff_0_HgTe_E0}, the above Hamiltonian has three extra IB terms $D_{1,2,3}$ brought by the electric field.
In fact, it is now in the most general symmetry-allowed form up to the second order of the momentum for the HgTe/CdTe QW along the (111) direction.
In addition, the parameter $m$ (mass term) can also be controlled by electric field.
In the contrast to (001) QW, the constant ($k$-independent) IB terms in \refcite{Konig2007QSHHgTe} are forbidden in \eqnref{eq:heff_0_HgTe} by the $C_3$ symmetry.
The $\mathcal{E}$ dependence of the parameters are shown in \tabref{tab:para_val_eff} for $d=62\AA$.
Since \eqnref{eq:heff_1_HgTe} is in the most general form, the electron-strain coupling for $\mathcal{E}\neq 0$ still keeps the form of \eqnref{eq:heff_1_HgTe}.
With \eqnref{eq:heff_0_HgTe}, \eqnref{eq:heff_1_HgTe}, the parameter expression, and $\xi_{1,2,3,4}=1$eV comparable as those in \refcite{Rostami2018PiE}, the PET jump can be calculated.

\subsection{Projection of the Kane Model}

With bases  $(\ket{\Gamma_6,\frac{1}{2}}$, $\ket{\Gamma_6,-\frac{1}{2}}$, $\ket{\Gamma_8,\frac{3}{2}}$, $\ket{\Gamma_8,\frac{1}{2}}$, $\ket{\Gamma_8,-\frac{1}{2}}$, $\ket{\Gamma_8,-\frac{3}{2}})$, the 6-band Kane model that we use for the (111) quantum well without the electric field reads
\eq{
\label{eq:Kane}
h_{Kane}(\bsl{k})=
\mat{
h_{\Gamma_6}(\bsl{k}) & T(\bsl{k}) \\
T^\dagger(\bsl{k}) & h_{\Gamma_8}(\bsl{k})
}\ ,
}
where $\bsl{k}=(k_1,k_2,k_3)$ with $k_3=-\ii \partial_{x_3}$, $h_{\Gamma_6}(\bsl{k})=\left(E_c+\frac{\hbar^2}{2 m_0}\left[ (2F+1)(k_1^2+k_2^2)+k_3(2F+1)k_3\right]\right)\sigma_0$,
\eq{
T(\bsl{k})=
\left(
\begin{array}{cccc}
 -\frac{1}{\sqrt{2}}k_+ P & \sqrt{\frac{2}{3}} k_{3} P & \frac{1}{\sqrt{6}}k_- P & 0 \\
 0 & -\frac{1}{\sqrt{6}}k_+ P & \sqrt{\frac{2}{3}} k_{3} P & \frac{1}{\sqrt{2}}k_-P \\
\end{array}
\right)\ ,
}
\eq{
h_{\Gamma_8}(\bsl{k})= \left(
\begin{array}{cccc}
 U+V & W & \widetilde{W} & 0 \\
 W^\dagger & U-V & 0 & \widetilde{W} \\
\widetilde{W}^\dagger & 0 & U-V & -W \\
 0 & \widetilde{W}^\dagger & -W^\dagger & U+V \\
\end{array}
\right)\ ,
}
$U=E_v-\frac{\hbar^2}{2 m_0}[(k_1^2+k_2^2)\gamma_1+k_3 \gamma_1 k_3]$, $V=\frac{\hbar^2}{2 m_0}[-(k_1^2+k_2^2)\gamma_3+2 k_3 \gamma_3 k_3]$, $W=\frac{1}{\sqrt{3}}\frac{\hbar^2}{2 m_0}[-\ii\sqrt{2} k_+^2 (\gamma_2-\gamma_3)+k_- \{k_3,2\gamma_2+\gamma_3\}]$, $\widetilde{W}=\frac{1}{\sqrt{3}}\frac{\hbar^2}{2 m_0}[ k_-^2 (\gamma_2+2\gamma_3)-\ii\sqrt{2}k_+\{k_3,\gamma_2-\gamma_3\}]$, $m_0$ is the mass of the electron, and the IB effect is neglected.
The electric field can be included by adding
\eq{
\label{eq:V_e}
V_e=-e\mathcal{E}x_3\mathds{1}_6
}
to \eqnref{eq:Kane}.

Due to the spatial dependence of the parameters, we require the anti-commutation form of some $k_3$-dependent terms, such as $\{k_3,\gamma_2\}$, to keep the Hamiltonian hermitian~\cite{Winkler2003SOC}.
The quantum well considered has the structure Hg$_{0.3}$Cd$_{0.7}$Te/HgTe/Hg$_{0.3}$Cd$_{0.7}$Te, leading to the $x_3$ dependence of parameters $X=E_{v,c}, F, \gamma_{1,2,3}$ as
\eq{
X=
\left\{
\begin{array}{l}
X^{A},\ |x_3|<\frac{d}{2}\\
X^{B},\ |x_3|>\frac{d}{2}
\end{array}
\right.\ .
}
The numerical values of the parameters in \eqnref{eq:Kane} are listed in \tabref{tab:Kane_para}.

\begin{table}[t]
    \centering
    \begin{tabular}{|c|c|c|c|c|c|c|}
    \hline
     $E_v ^{A}$ & $E_c^{A} $ & $ F^{A} $ & $ \gamma_1^{A} $ & $ \gamma_2^{A} $ & $ \gamma_3^{A} $  & $P$\\
    \hline
    $0$ & $-0.303$ & $0$ & $4.1$ & $0.5$ & $1.3$ & $8.47$\\
    \hline
    $E_v ^{B}$ & $ E_c^{B} $ & $ F^{B} $ & $ \gamma_1^{B} $ & $ \gamma_2^{B} $ & $ \gamma_3^{B} $ & \\
    \hline
    $-0.399$ & $0.607$ & $-0.063$ & $2.26$ & $-0.046$ & $0.411$ &\\
    \hline
    \end{tabular}
\caption{Values of parameters in \eqnref{eq:Kane} for Hg$_{0.3}$Cd$_{0.7}$Te/HgTe/Hg$_{0.3}$Cd$_{0.7}$Te quantum well.
The unites of $E_v, E_c$ are eV, the unit of $P$ is eV$\AA$, and other parameters are dimensionless~\cite{Laurenti1990HgCdTe,Novik2005HgMnTe}.
}
\label{tab:Kane_para}
\end{table}

The effective models are derived according to \refcite{Bernevig2006BHZ,Rothe2010HgTe}.
We first numerically obtain the wavefunctions of E1, H1, LH1, HH2, and HH3 bands at $k_1=k_2=\mathcal{E}=0$, and project the remaining terms to the bases to get a $10\times 10$ Hamiltonian.
Then, we project the $10\times 10$ Hamiltonian to the E1 and H1 bands with second order perturbation to get \eqnref{eq:heff_0_HgTe_E0} and \eqnref{eq:heff_0_HgTe}.
Keeping terms up to $k^2$ and $\mathcal{E}^2$ order, the values of the parameters are listed in \tabref{tab:para_val_eff_E0} and \tabref{tab:para_val_eff}.

\begin{table*}[t]
\centering
\begin{tabular}{ccccccc}
\hline
$d/\AA$ & $m_0$/eV & $B_0$/(eV \AA$^2$) & $m$/eV & $B$/(eV \AA$^2$) & $A_1$/(eV \AA) & $A_2$/(eV \AA) \\
\hline\hline
 60.00 & -0.006700 & 39.88 & 0.005600 & 60.45 & 3.595 & 0.1248 \\
 61.00 & -0.007570 & 40.90 & 0.004370 & 61.46 & 3.582 & 0.1237 \\
 62.00 & -0.008400 & 41.94 & 0.003200 & 62.51 & 3.569 & 0.1225 \\
 63.00 & -0.009240 & 42.99 & 0.002040 & 63.56 & 3.555 & 0.1214 \\
 64.00 & -0.01009 & 44.08 & 0.0008850 & 64.65 & 3.540 & 0.1204 \\
 65.00 & -0.01084 & 45.14 & -0.0001650 & 65.71 & 3.527 & 0.1193 \\
 66.00 & -0.01159 & 46.25 & -0.001210 & 66.82 & 3.514 & 0.1182 \\
 67.00 & -0.01235 & 47.41 & -0.002250 & 67.98 & 3.500 & 0.1172 \\
 68.00 & -0.01307 & 48.54 & -0.003230 & 69.12 & 3.486 & 0.1162 \\
 69.00 & -0.01374 & 49.75 & -0.004160 & 70.33 & 3.473 & 0.1152 \\
 70.00 & -0.01442 & 50.98 & -0.005080 & 71.56 & 3.459 & 0.1143 \\
 \hline
\end{tabular}
\caption{Parameter values for \eqnref{eq:heff_0_HgTe_E0} at various widths $d$.}
\label{tab:para_val_eff_E0}
\end{table*}

\begin{table*}[t]
\centering
\begin{tabular}{|c|c|c|}
\hline
 $m_0$/eV & $B_0$/(eV \AA$^2$) & $m$/eV \\
 \hline
 4182 $(-e\mathcal{E} /(\text{eV}\AA^{-1}))^2-$0.008400 & 544800 $(-e\mathcal{E} /(\text{eV}\AA^{-1}))^2$+41.94 & 0.003200 -17.21 $(-e\mathcal{E} /(\text{eV}\AA^{-1}))^2$\\
\hline\hline
 $B$/(\text{eV} \AA$^2$) & $A_1$/(\text{eV} \AA) & $A_2$/(\text{eV} \AA)\\
\hline
 534600 $(-e\mathcal{E} /(\text{eV}\AA^{-1}))^2$+62.51 & 67320 $(-e\mathcal{E} /(\text{eV}\AA^{-1}))^2$+3.569 & 293.0 $(-e\mathcal{E} /(\text{eV}\AA^{-1}))^2$+0.1225 \\
\hline\hline
  $D_1$/(\text{eV} \AA$^2$) & $D_2$/(\text{eV} \AA$^2$) & $D_3$/(\text{eV} \AA)\\
\hline
724.7 $(-e\mathcal{E} /(\text{eV}\AA^{-1}))$ & -1947 $(-e\mathcal{E} /(\text{eV}\AA^{-1}))$ & -1196 $(-e\mathcal{E} /(\text{eV}\AA^{-1}))$\\
\hline
\end{tabular}
\caption{Parameter values for \eqnref{eq:heff_0_HgTe} for $d=62\AA$.}
\label{tab:para_val_eff}
\end{table*}

\subsection{Construction of the Hamiltonian based on symmetry}

As discussed in the main text, the symmetry group of interest is generated by  the three-fold rotation $C_3$ along $(111)$, and the mirror $m_{1\bar{1}0}$ perpendicular to $(1,\bar{1},0)$ and the TR operation $\mathcal{T}$.
With the bases $(\ket{E_1,+},\ket{H_1,+},\ket{E_1,-},\ket{H_1,-})$, those symmetry operations,  according to the convention in \refcite{Winkler2003SOC}, are represented as
\eqa{
C_3&\dot{=}\left(
\begin{array}{cccc}
 \ee^{-\frac{\ii \pi}{3}} & 0 & 0 & 0 \\
 0 & -1 & 0 & 0 \\
 0 & 0 & \ee^{\frac{\ii \pi }{3}} & 0 \\
 0 & 0 & 0 & -1 \\
\end{array}
\right)\\
m_{1\bar{1}0}&\dot{=}-\ii\sigma_x\tau_0=\left(
\begin{array}{cccc}
 0 & 0 & -\ii & 0 \\
 0 & 0 & 0 & -\ii \\
 -\ii & 0 & 0 & 0 \\
 0 & -\ii & 0 & 0 \\
\end{array}
\right)\\
\mathcal{T}&\dot{=}-\ii\sigma_y\tau_0=\left(
\begin{array}{cccc}
 0 & 0 & -1 & 0 \\
 0 & 0 & 0 & -1 \\
 1 & 0 & 0 & 0 \\
 0 & 1 & 0 & 0 \\
\end{array}
\right)\cc \ ,
}
where $\tau$'s and $\sigma$'s are Pauli matrices for $E_1,H_1$ indexes and $\pm$ indexes, respectively.
According to the symmetry representations, the matrix and momenta of the effective model can be classified as \tabref{tab:HgTe_Sym}.

\begin{table*}[t]
    \centering
    \begin{tabular}{cc}
    \hline
    IR & Expressions\\
    \hline
    \hline
         $A_1,+$ &  $\sigma_0\tau_0$, $\sigma_0\tau_z$, $k_1^2+k_2^2$, $u_{11}+u_{22}$  \\
         $A_1,-$ &  $\sigma_x\tau_-$  \\
         $A_2,+$ &\\
         $A_2,-$ &  $\sigma_y\tau_-$, $\sigma_z\tau_0$, $\sigma_z\tau_z$\\
         $E,+$ & $(\sigma_z\tau_y,\sigma_0\tau_x)$, $(\sigma_y\tau_y,\sigma_x\tau_y)$, $(2 k_1 k_2,k_1^2-k_2^2)$, $(u_{12}+u_{21},u_{11}-u_{22})$\\
         $E,-$ & $(\sigma_z\tau_x,-\sigma_0\tau_y)$,$(\sigma_y\tau_x,\sigma_x\tau_x)$,$(\sigma_y\tau_+,-\sigma_x\tau_+)$, $(k_1,k_2)$\\
         \hline
    \end{tabular}
    \caption{The irreducible representations (IRs) of $C_{3v}$ and TR symmetries.
    $A_1$, $A_2$ and $E$ are IRs of $C_{3v}$ and $\pm$ are parity under TR.
    $\tau_\pm=(\tau_0\pm \tau_z)/2$.
    }
    \label{tab:HgTe_Sym}
\end{table*}
From \tabref{tab:HgTe_Sym}, the most general symmetry-allowed Hamiltonian without the electron-strain coupling can be derived to the $k^2$ order, resulting in \eqnref{eq:heff_0_HgTe}.
As shown in \tabref{tab:HgTe_Sym}, $u_{ij}$ behaves the same as the $k^2$ term, and thereby the electron-strain coupling has the same form as the $k^2$ term in \eqnref{eq:heff_0_HgTe}.

\section{BaMnSb$_2$}
\label{app:bms}

In this section, we include more details on BaMnSb$_2$.

\begin{figure}[t]
\includegraphics[width=0.5\columnwidth]{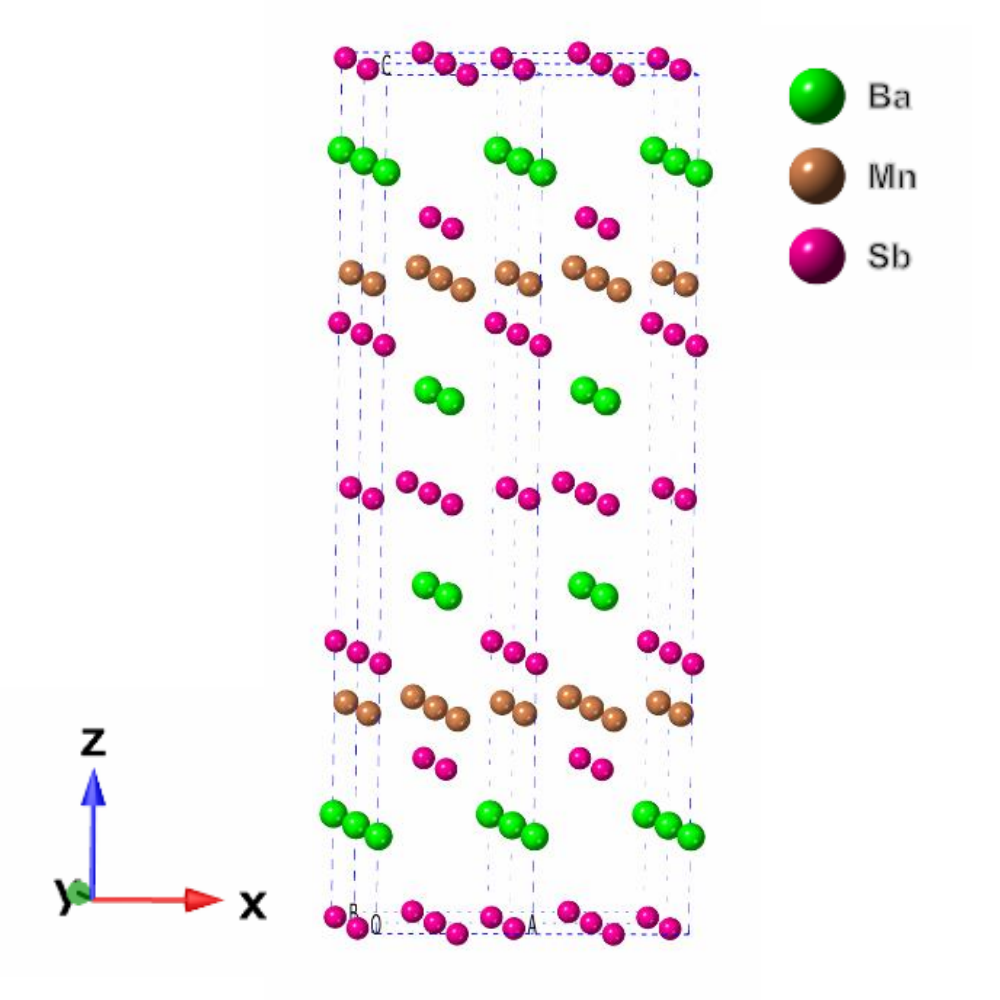}
\caption{
The crystalline structure of BaMnSb$_2$ generated from CrystalMaker.
}
\label{fig:BMS_structure}
\end{figure}

\subsection{Review}
In this part, we review the form and the dispersion of the TB model derived in \refcite{Liu2019SCM3DQHE} for integrity.
This part does not contain any original results.
More details can be found in \refcite{Liu2019SCM3DQHE}.

According to the main text, there are two Sb atoms in one unit cell, labeled as 1 and 2, that have sub-lattice vectors $\boldsymbol{\tau}_1=(x_1 a,0)$ and $\boldsymbol{\tau}_2=(x_2 a,b/2)$, respectively.
$a,b$ are the lattice constants of the unit cell in $x,y$ direction and the values of $x_{1,2}$ are given later.
Combined with $p_x$ and $p_y$ orbitals, the bases of the TB model are $\ket{\bsl{R}+\boldsymbol{\tau}_i,\alpha,s}$ with the lattice vector $\bsl{R}=(l_x a, l_y b)$ ($l_{x,y}\in \mathds{Z}$), the sublattice index $i=1,2$, the orbital index $\alpha=p_x,p_y$, and the spin-$z$ index $s=\uparrow,\downarrow$.
The TB model consists of the on-site term $H_0$, the nearest-neighboring (NN) hopping $H_1$ and the next-NN hopping $H_2$ in the TB model, \ie $H_{TB}=H_0+H_1+H_2$.
$H_0$ has the form
\eq{
\label{eq:H0}
H_0=\sum_{\bsl{R},i}c^\dagger_{\bsl{R}+\boldsymbol{\tau}_i}M_i c_{\bsl{R}+\boldsymbol{\tau}_i}
}
with
\eq{
c^\dagger_{\bsl{R}+\boldsymbol{\tau}_i}=(c^\dagger_{\bsl{R}+\boldsymbol{\tau}_i,p_x,\uparrow},c^\dagger_{\bsl{R}+\boldsymbol{\tau}_i,p_x,\downarrow},c^\dagger_{\bsl{R}+\boldsymbol{\tau}_i,p_y,\uparrow},c^\dagger_{\bsl{R}+\boldsymbol{\tau}_i,p_y,\downarrow})\ .
}
$H_1$ reads
\eq{
\label{eq:H1}
H_1=\sum_{\bsl{R}}\sum_{n=1}^4c^\dagger_{\bsl{R}+\Delta\bsl{R}_n+\boldsymbol{\tau}_2}T_n c_{\bsl{R}+\boldsymbol{\tau}_1}+h.c.\ ,
}
where $\Delta\bsl{R}_1=(0,0)$, $\Delta\bsl{R}_2=(a,0)$, $\Delta\bsl{R}_3=(a,-b)$ and $\Delta\bsl{R}_4=(0,-b)$.
$H_2$ reads
\eq{
\label{eq:H2}
H_2=\sum_{\bsl{R},i}\sum_{n=x,y}c^\dagger_{\bsl{R}+\Delta\bsl{R}_n+\boldsymbol{\tau}_i}Q_{ni} c_{\bsl{R}+\boldsymbol{\tau}_i}+h.c.\ ,
}
where $\Delta\bsl{R}_x=(a,0)$ and $\Delta\bsl{R}_y=(0,b)$.
The forms of $M$'s, $T$'s, and $Q$'s are
\eqa{
& M_1=\widetilde{m}_0\tau_0\sigma_0+\widetilde{m}_1\tau_z\sigma_0+\lambda_0\tau_y\sigma_z\ ,\\
&M_2=C^{OS}_{4z}M_1 (C^{OS}_{4z})^\dagger\ ,\\
&T_1=t_0\tau_0\sigma_0+t_1\tau_x\sigma_0+i t_2\tau_y\sigma\ ,T_{2}=\frac{\tau_z\sigma_y T_{1} \tau_z\sigma_y}{f(\alpha)}\ ,\\
&T_{4}=\tau_z\sigma_y T_{1} \tau_z\sigma_y\ ,\ T_{3}=\frac{T_{1}}{f(\alpha)}\ ,\\
& Q_{x1}=t_3 \tau_0\sigma_0+t_4\tau_z\sigma_0\ ,\ Q_{x2}=t_5\tau_0\sigma_0+t_6\tau_z\sigma_0\ ,\\
&Q_{y1}=C^{OS}_{4z}Q_{x2} (C^{OS}_{4z})^\dagger\ ,\ Q_{y2}=C^{OS}_{4z}Q_{x1} (C^{OS}_{4z})^\dagger\ ,
}
where $f(\alpha)=0.2 \alpha +1$, and $C_{4z}^{OS}=(-\ii \tau_y) \ee^{-\ii \frac{\sigma_z}{2}\frac{\pi}{2}}$ is the representation of the four-fold rotation along $z$ in the orbital and spin subspace.
$\alpha$ is the dimensionless distortion parameter; $\alpha=0$ and $\alpha=1$ correspond to the non-distorted and fully distorted cases, respectively.
Moreover, the distortion effect on the relative atom positions is chosen as $x_1=\frac{1}{2}+(0.4512-\frac{1}{2})\alpha$ and $x_2=0.01729\alpha$, while we neglect distortion-induced change of $a$ and $b$.

The numerical calculation is done for
\eqa{
\label{eq:para_I4mmm}
&\widetilde{m}_{0}=0\ ,\ \widetilde{m}_{1}= 0.3\text{eV}\ ,\ \lambda_{0}=0.25eV\ ,\ t_{0}= 1\text{eV}\ ,\\
& t_{1}= 2\text{eV}\ ,\ t_{2}= 0\ ,\ t_{3}= 0.1\text{eV}\ , t_{4}= -0.06\text{eV}\ ,\\
&t_{5}= 0.15\text{eV}\ ,\ t_{6}= -0.06\text{eV}\ ,\ \text{and }a=b=4.5\AA\ .
}
The energy dispersion for $\alpha=1$ is shown in the supplementary material of \refcite{Liu2019SCM3DQHE}.

\subsection{TB Calculation of PET}

The main effect of the strain in the TB model is to change the hopping amplitudes among atoms~\cite{Li2016StrainTMD,Wang2018TBPiezo}, which can be modeled by performing the following replacement~\cite{Li2016StrainTMD} to the hopping parameters:
\eq{
t_{ab}\rightarrow \left(1- \beta \frac{\delta_{i}\delta_{j}u_{ij}}{|\boldsymbol{\delta}|^2}\right)t_{ab}
}
, where $t_{ab}$ is the hopping parameter between atoms at $\bsl{r}_a$ and $\bsl{r}_b$ in the non-deformed case, and $\boldsymbol{\delta}=\bsl{r}_a-\bsl{r}_b$.
$\beta$ is the electron-phonon coupling parameter whose value for BaMnSb$_2$ has not been determined, and thereby we adopt the typical value $\beta=2$ for the TMDs~\cite{Li2016StrainTMD} to give a reasonable estimation of the PET jump.

\subsection{Effective Model Analysis}
To analytically demonstrate the PET jump, we project the tight-binding model into the subspace spanned by two degenerate states at each gap closing point (valley).
As discussed in \refcite{Liu2019SCM3DQHE}, the resultant effective model reads
\eq{
\label{eq:h_q_BMS}
h_{\pm}^{(0)}(\bsl{q})=(E_0\pm v_0 q_y)\tau_0\pm v_2 q_y \tau_z\pm v_1 q_x \tau_x\pm (E_1+\lambda)\tau_y\ ,
}
where $h_\pm(\bsl{q})$ is around $\bsl{K}_\pm$ with $\bsl{q}=\bsl{k}-\bsl{K}_\pm$, the two bases of $h_+^{(0)}$ ($h_-^{(0)}$) are $\ket{\bsl{K}_+, p_x\pm \ii p_y, \uparrow}$ ($\ket{\bsl{K}_-, p_x\pm \ii p_y, \downarrow}$), and the term with small coefficient has been omitted.
$E_1$ and $v_1$ are given by the distortion,  $\lambda$ labels the SOC strength, and we choose $E_1 < 0,\lambda > 0$ without loss of generality.
According to \eqnref{eq:h_q_BMS}, the gap closing can be achieved by tuning the distortion parameter $E_1$ to $E_1+\lambda=0$, which changes the $Z_2$ index since only one Dirac cone appears in half 1BZ.
To study the PET jump, we include the electron-strain coupling with the form
\eq{
\label{eq:h_str_BMS}
h_{\pm}^{(1)}(\bsl{q})=N_0 \tau_0+N_1 \tau_x+N_2 \tau_z\ ,
}
where $N_1=\xi_{xy} (u_{xy}+u_{yx})$ and $N_i=\xi_{i,xx}u_{xx}+\xi_{i,yy} u_{yy} $ for $i=0,2$.
It is derived from the symmetry consideration and the fact that the $\tau_y$ term is valley-dependent and thus of higher order.
Combining the above equation with \eqnref{eq:h_q_BMS}, we obtain the non-zero PET jump
\eqa{
&\Delta\gamma_{xxx}=-e\frac{\sgn{v_1 v_2}}{\pi v_2 }\xi_{2,xx}\\
&\Delta\gamma_{xyy}=-e\frac{\sgn{v_1 v_2}}{\pi v_2 }\xi_{2,yy}\\
&\Delta\gamma_{yxy}=\Delta\gamma_{yyx}=e\frac{ \sgn{v_1 v_2} }{\pi v_1 }\xi_{xy}\ ,
}
as $E_1$ is tuned from $-\lambda+0^-$ to $-\lambda+0^+$.
Therefore, the gap closing and the PET jump are consistent with result for scenario (iii) of $p1m1$.

\bibliography{bibfile_references}

\end{document}